\documentclass[12pt]{article}

\usepackage{newtxtext,newtxmath}

\usepackage{graphicx}
\usepackage{tabularx}
\usepackage{booktabs}
\usepackage[letterpaper,margin=1in]{geometry}

\renewenvironment{abstract}
	{\quotation}
	{\endquotation}

\date{}

\makeatletter
\renewcommand{\fnum@figure}{\textbf{Figure \thefigure}}
\renewcommand{\fnum@table}{\textbf{Table \thetable}}
\makeatother

\usepackage{scicite}

\usepackage{url}

\def\scititle{
	A thermoinformational formulation for the description of neuropsychological systems
}
\title{\bfseries \boldmath \scititle}

\author{
    George-Rafael Domenikos$^{1,*}$,
	Victoria Leong$^{1}$,
    \and
	\small$^{1}$Department of Social Sciences, EMPOWER Centre, Nanyang Technological University, 308232 Singapore\and
	\small$^\ast$Corresponding author. Email: georgios.rd@ntu.edu.sg\and
}

\begin{document} 

\maketitle

\begin{abstract} \bfseries \boldmath
Complex systems produce high-dimensional signals that lack macroscopic variables analogous to entropy, temperature, or free energy. This work introduces a thermoinformational formulation that derives entropy, internal energy, temperature, and Helmholtz free energy directly from empirical microstate distributions of arbitrary datasets. The approach provides a data-driven description of how a system reorganizes, exchanges information, and moves between stable and unstable states.
Applied to dual-EEG recordings from mother–infant dyads performing the A-not-B task, the formulation captures increases in informational heat during switches and errors, and reveals that correct choices arise from more stable, low-temperature states. In an independent optogenetic dam-pup experiment, the same variables separate stimulation conditions and trace coherent trajectories in thermodynamic state space.
Across both human and rodent systems, this thermoinformational formulation yields compact and physically interpretable macroscopic variables that generalize across species, modalities, and experimental paradigms.

\end{abstract}

\noindent

\section{Introduction}

Neural and behavioral systems exhibit extraordinary complexity, challenging our ability to describe and predict their dynamics. This situation is reminiscent of early physics, when scientists characterized the macroscopic behavior of gases long before deciphering molecular details. The solution then was thermodynamics, a framework that quantified observable properties (pressure, temperature, entropy; here ‘temperature’ denotes the inverse sensitivity of entropy to internal energy) and later bridged to microscopic theory via statistical mechanics. By analogy, a thermodynamic perspective may help in taming some of the complexity of brain and behavior, providing macroscopic laws for neural systems that do not require complete microscopic knowledge. Indeed, the brain is fundamentally a physical information-processing device, suggesting that principles from thermodynamics and information theory can be jointly applied to understand neural activity \cite{collell2015brain, parr2019generalised, seifert2012stochastic}. Researchers have long noted deep theoretical links between energy and information: Landauer’s limit formalized the minimal energy cost of erasing a bit of information \cite{landauer1996minimal}, while Schrödinger famously invoked “negative entropy” to explain how living systems sustain order \cite{rio2011thermodynamic,england2015dissipative}.

Recent theoretical frameworks have begun to merge brain science with thermodynamics in earnest. Notably, \textbf{Friston’s Free-Energy Principle} (FEP) models the brain as a thermodynamic, like system that maintains its organization by minimizing “free energy,” an information-theoretic proxy for surprise or prediction error \cite{friston2010free,friston2023free}. Although variational free energy is formally the difference between an expected energy term and an entropy term, it is an information theoretic bound on model evidence, not a thermodynamic potential. Minimising the free energy drives an agent to states that are unsurprising under its generative model, whereas minimising the physical Helmholtz free energy moves a system toward thermodynamic equilibrium and can proceed through either entropy increase (e.g. salt dissolving) or decrease, depending on boundary conditions. Consequently, physical free-energy dissipation does not imply monotonic entropy decline; a system can lower $F_H$ by releasing internal energy into the bath while its entropy rises. This distinction implies that stability in an adaptive system may emerge either through the reduction of free energy or through controlled increases in entropy and temperature that maintain a robust repertoire under changing conditions.

Complementary approaches emphasize the opposite but related mandate: rather than minimizing an information bound free energy, neural circuits may maximize entropy production in their physical processes. Varpula et al. \cite{varpula2013thoughts} showed that neural flows tend to follow paths that dissipate free energy as rapidly as possible, potentially explaining habits, memory consolidation, and other repetitive behaviors as energetically favored trajectories. Despite the differing emphases of informational free-energy minimization versus physical free-energy dissipation, both perspectives underscore that brain function obeys organizing principles drawn from thermodynamics and information theory.

Mounting empirical evidence supports the idea that the brain operates at the nexus of thermodynamics and information. Neuroscientists have applied entropy and related information metrics to brain signals, revealing insightful patterns. The “entropic brain” hypothesis \cite{carhart2018entropic} proposes that the richness of conscious states correlates with higher entropy in neural activity, a notion supported by EEG, MEG, and fMRI studies in psychedelic, wakeful, anesthetized, and aging brains. Generally, lower entropy in neural time series is linked to diminished consciousness and cognitive impairment, whereas healthy cognition appears to balance order and flexibility \cite{huang2020temporal}. Indeed, many studies suggest that the cortex operates near a critical point a transitional state between order and chaos that optimizes information processing. Spontaneous cortical activity exhibits scale invariant neuronal avalanches, following power-law size distributions \cite{beggs2003neuronal}, and resting state networks emerge naturally when connectome based models are tuned to this critical regime \cite{haimovici2013brain}. Such observations make concepts like temperature, phase transitions, and order parameters quantitatively concrete in neuroscience. Crucially, new technologies now enable direct tests of these physics-inspired principles. Optogenetics acts as an experimental “thermostat,” perturbing circuits with millisecond precision. Studies using optogenetic stimulation have revealed abrupt, all or none transitions that herald epileptic seizures, demonstrating phase-transition like tipping points in vivo. Controlled light driven synchronization further allows researchers to track trajectories through neural state spaces, observing how ensembles migrate between regimes of relative disorder and organized coordination. Collectively, a convergence of theory and evidence now points toward treating neural and cognitive phenomena within the formalism of thermodynamics and statistical physics. In this context, entropy reflects the breadth of accessible neural configurations; increases in entropy indicate a richer repertoire of coordinated microstates that support adaptive and flexible behavior, rather than mere disorder.

Beyond neuroscience, thermoinformational modeling has already proven fruitful in social and biological collectives, underscoring its generality. For instance, Domenikos and Mantzaris developed an empirical formulation akin to a thermodynamic equation of state for electoral opinion dynamics, mapping political “temperature,” “pressure,” and “volume” and revealing isothermal and isentropic cycles analogous to ideal-gas behavior \cite{domenikos2024possible}. A prior study had already framed political polarization as an isolated thermodynamic system, whose entropy trace predicts opinion space segregation over time \cite{mantzaris2023exploring}. Earlier work on voter model simulations quantified political segmentation directly through entropy estimation \cite{domenikos2022model}. Extending into biology, the same authors showed that an agent-based bird-flocking simulation can be characterized by entropy, internal energy, temperature, and heat, with flocks undergoing cooling-like trends as they condense in flight \cite{mantzaris2025exploring}. These cross-domain successes suggest that thermodynamic variables provide unifying, interpretable descriptors for emergent organization in complex systems, setting the foundation for the full thermoinformational modeling of data systems.

Motivated by this progress, we introduce a generalized thermoinformational framework for neural and behavioral systems. This method aims to formalize statistical physics tools to arbitrary data streams from brain and behavior, translating time-series measurements into ensembles of system states from which entropy, temperature, free energy, and related variables can be computed. It is validated in two contrasting paradigms: (i) \textbf{EEG recordings} from mother–infant dyads performing the standard A-not-B task, and (ii) \textbf{optogenetically driven synchrony} in rodent prefrontal circuits. In the first paradigm, thermoinformational measures capture systematic shifts in neural entropy and other state variables that predict task outcomes directly from raw EEG, revealing how the infant brain reorganizes between memory representations (“A” vs. “B” choices). In the second, the same variables trace distinct state-space trajectories under different light stimulation protocols, delineating separations in the network’s thermodynamic state as it entrains or diverges. By bridging cognitive-level behavior and circuit level dynamics within one quantitative framework, these applications illustrate the versatility of thermoinformational analysis. More broadly, they demonstrate how treating brain activity as a thermodynamic process can yield measurable insights, from identifying reconfiguration events through heat flux to characterizing robust high entropy regimes preceding correct decisions. Such a unified thermo-information framework aims to advance the understanding of neuropsychological systems by providing common, physically grounded descriptors for complex neural and behavioral phenomena, just as classical thermodynamics once unified heat, work, and matter. The results highlight the promise of this approach for uncovering fundamental state variables of the brain and point toward a new, integrative paradigm for quantifying mind and life in terms of energy and information.

\section{Materials and Methods}

\subsection{Mathematical Model}
Most prior applications of thermodynamic concepts to complex systems have focused on individual variables or limited subsets (e.g., entropy or free energy) within specific contexts. Here, this approach is extended by deriving a complete and internally consistent set of thermodynamic variables, entropy, internal energy, temperature, specific heat, and free energy—and linking them through an empirical relation of key variables, akin to an equation of state. This unified framework enables quantitative description of data systems outside traditional physics, providing a physically grounded basis for modeling and interpretation across diverse domains.

\subsubsection{Data System and Microstate Definition}

To describe any data system thermoinformationally, the microstate ought to be first defined as the smallest measurable unit of the system.  
In physical systems, this corresponds to the energy of a particle; in abstract domains, it may represent the value of a single measured element (e.g., a stock price, behavioral score, or EEG feature).  
To generalize beyond physics, these microstates are treated as \emph{vectors} rather than scalars, enabling multidimensional descriptions such as behavioral or neural configurations at each observation point.

Formally, a microstate at index \(t\) is
\begin{equation}
\label{eq:Microstate}
    \vec{X}^{\,t} = (X_1^t, X_2^t, \dots, X_N^t),
\end{equation}
where \(X_i^t\) are the attributes measured at that index.  
Measurements need not represent time per se, they may correspond to locations, modalities, or other indices depending on the system.

For a dataset containing multiple individuals, measurements, and sampling points, the complete data system can be expressed as
\begin{equation}
\label{mat:system_matrix}
    \vec{A} = \vec{X}\{t, n, i\},
\end{equation}
with \(t = 0\dots\tau\) denoting temporal (or spatial) indices, \(n = 1\dots N\) the measurement types, and \(i = 1\dots I\) the individual entities (e.g., subjects, animals, or assets).  
The probability distribution \(p(E)\) is constructed from the ensemble of microstates defined over this multidimensional matrix.

This framework generalizes naturally to higher-dimensional data, where
\[
\vec{A} = \vec{X}\{a,b,\dots,\omega\}, \quad \vec{A} \in \mathbb{R}^{a\times b\times \dots \omega},
\]
and the microstate \(\vec{x} \in \mathbb{R}^{(a\times b\times \dots \omega)-1}\) encodes the desired informational unit.  
The chosen dimensionality of \(\vec{x}\) determines which aspects of the system’s organization, temporal, spatial, or relational, are captured.  
A schematic example of the data matrix structure is provided in Supplementary Section~S1.

\subsubsection{Derivation of Thermoinformational variables}
The proposed thermoinformational framework is set in a generalized form in order to be able to be applied to different types of datasets and dimensionalities.

Assume microstate $E$ from system $\vec{A}$. If $E$ is scalar, then $|E|=E$, which is the microscopical energy for the microstate $E$. If $E\in R^z$, with $z>1$, then the energy of the microstate is the $L2$ norm of the vector.
 
In order to begin setting up the thermodynamic variables, we need to establish the distribution function of the microstates. In classical thermodynamics, the distribution function depends on the nature of the particles (like Maxwell-Boltzmann for classical systems, Bose-Einstein for bosonic systems, etc.). Based on this distribution function, the entropy and temperature are defined.
 
In arbitrary data systems, the distribution function of the data is not standard, so the temperature and entropy need to be re-derived based on an unknown distribution.
 
To start this process, the distribution function needs to be established. For simplicity, the derivation will be performed here for a scalar microstate of an unknown distribution.  
 
The probability of finding each microstate is calculated as:

\begin{equation}
    \label{eq:microstate_prob}
    p(E_{k})= \frac{\sum_{i=1}^N \delta_{E_{k},E_i}}{N}
\end{equation}

with $p(E_{k})$ being the probability of finding the microstate $E_k$, with $k=1\dots D$ and $D$ being the number of distinct microstates. $N$ is the total number of microstates, and $\delta$ is Kronecker's delta.
For the continuation of the calculations, it is beneficial to have the probability distribution in the form of a continuous (Riemann differentiable) function. The method can be also applied in discretized systems. 
 
To find the probability distribution function, any numerical fitting method can be used, provided that it reaches sufficient accuracy. Typical such methods include: Kernel Density Estimation, Regression models, Dirichlet process, Gaussian Mixture Models, etc.
 
Utilizing the most suitable method for each system, we gain the continuous and Riemann differentiable function $p(E)$, with $E\in [E_{min}, E_{max} ]$.

In informational systems the Shannon entropy is usually utilized and defined as:

\begin{equation}
    \label{eq:Shannon_Entropy}
    S = - \int_{E_{min}}^{E_{max}} p(E) ln(p(E)) dE
\end{equation}

Depending on the nature of the distribution of the microstates of the system, other definitions of the entropy might be more successful in capturing behaviors and phenomena. Other such entropies include the Tsallis, the Renyi, the Sharma-Mittal, the Bose-Einstein, etc. Any entropy definition that applies to the nature of the distribution of the dataset can be used.
 
The internal energy is defined analogously to its definition in statistical mechanics, meaning that it is defined as the expected value of the microstate.

\begin{equation}
    \label{eq:Internal_Energy}
    U=\int_{E_{min}}^{E_{max}} |E|\: p(E) dE
\end{equation}
In the case of a scalar microstate $|E|=E$.

The entropy and internal energy are the fundamental thermoinformational variables as they are calculated directly from the dataset. The second tier of thermoinformational variables to be defined are the ones that are still derived from the same microstate but involve differentiations of the fundamental variables, and are therefore more sensitive to the fit of the probability distribution function to the original data. These are the temperature, specific heat capacity, free energy (Helmholtz or Gibbs), and heat. 

The fundamental definition of the temperature in thermodynamics originates from the form:

\begin{equation}
    \label{eq:1st_law}
    \frac{1}{T}=\frac{\partial S}{\partial U}
\end{equation}

The temperature (T) and the internal energy (U) are key to the description of agent dynamics. Eq.~\ref{eq:1st_law} is a derivation of temperature based on the rate of change of entropy (S). This means that higher temperature corresponds to a smaller sensitivity of entropy to energy, that is, the system is already in a high-repertoire regime where additional energy yields little further increase in entropy.

In this case where $S(U)$ is known only numerically , one may approximate this derivative by introducing small energy perturbations and observing the corresponding entropy changes.

In practice, temperature was estimated by introducing infinitesimal perturbations to the kernel density estimate (KDE) of the microstate distribution rather than to the internal energy itself. Specifically, the KDE bandwidth $\lambda$ was scaled by a small factor $(1+\Delta\lambda)$, yielding a perturbed probability distribution $p_{\lambda+\Delta\lambda}(E)$. This controlled deformation of the distribution produced corresponding changes in the numerically computed entropy and internal energy, denoted $S_{\lambda+\Delta\lambda}$ and $U_{\lambda+\Delta\lambda}$. The derivative $\mathrm{d}S/\mathrm{d}U$ was then approximated as
\[
\frac{\mathrm{d}S}{\mathrm{d}U} \approx 
\frac{S_{\lambda+\Delta\lambda} - S_{\lambda}}{U_{\lambda+\Delta\lambda} - U_{\lambda}},
\]
and the generalized temperature was obtained from its reciprocal,
\[
T = \left(\frac{\mathrm{d}S}{\mathrm{d}U}\right)^{-1}.
\]
This procedure ensures that the perturbation remains internal to the estimated microstate distribution, maintaining thermodynamic consistency between $S$, $U$, and $T$, while capturing how small deformations in the inferred density translate into local energetic responsiveness of the system.

Similarly to the temperature the Specific Heat under constant volume (Cv) is defined as:

\begin{equation}
    \label{eq:Cv}
    Cv=T\frac{dS}{dT}
\end{equation}

which is to be solved similarly through perturbations as per the temperature calculations.

The free energy is straightforward to define: 

\begin{equation}
    \label{eq:Helmholtz}
    F_H=U-TS
\end{equation}
Lower free energy values are consistent with more efficient or stable configurations, although such stability can coexist with higher entropy or temperature when the system remains robust under changing inputs. The Helmholtz free energy is therefore a key quantity to compute: within this empirical framework, \(F_H(\rho,T)\) serves as the system’s fundamental thermodynamic equation (in the sense of a generating relation), from which the operational state relations follow by standard derivatives and Legendre transforms.\footnote{A fundamental thermodynamic relation is a generating relation that mathematically interconnects the primary state variables of a system, allowing other thermodynamic properties to be obtained by differentiation and Legendre transforms \cite{BlackHartley2010}} This usage is empirical and dataset-specific and does not assert a universal physical law for the brain. Here, temperature \(T\) is derived directly from the microstate distribution, while \(\rho\) denotes the effective microstate density used in the following applications.

\begin{figure}
     \centering
     \includegraphics[width=0.85\linewidth]{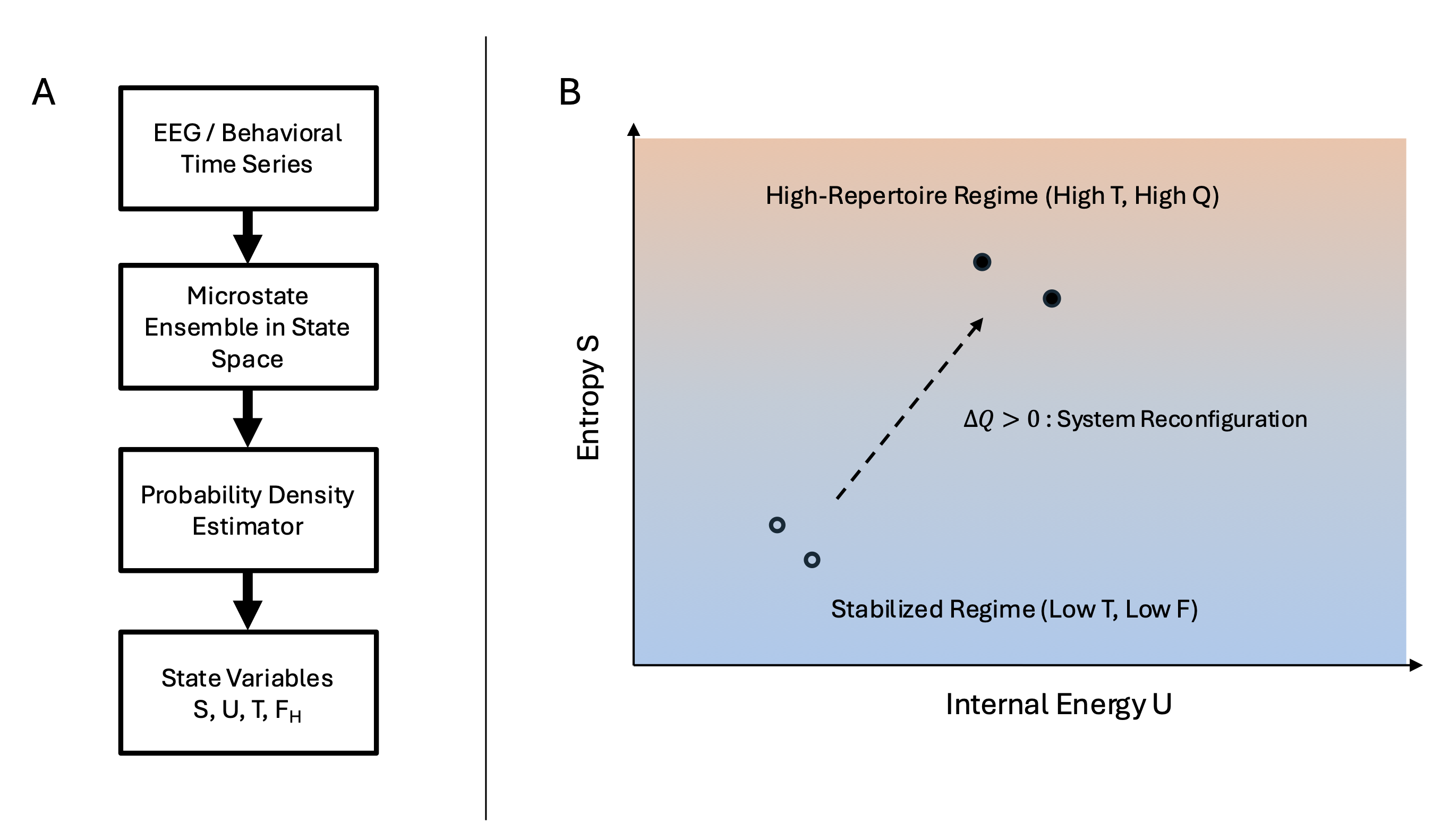}
     \caption{\textbf{(A)} Framework Flowchart. \textbf{(B)} \((S,U)\) state-space schematic with example displacements; \(\Delta Q>0\) = positive informational heat (reconfiguration). Illustrative regime labels (lower \(T,F_H\) vs.\ higher \(T,Q\)) aiding in general interpretability of framework's results.}
     \label{fig:Flow_Figure}
 \end{figure}

Fig. ~\ref{fig:Flow_Figure} summarizes the framework and its visualization. 
Panel~A shows the computational pipeline leading from an empirical dataset through a computational method to the thermoinformatics results. 
Panel~B illustrates how results can be plotted in a low-dimensional thermodynamic state space (here \((S,U)\)), where epochs or trials appear as points and changes appear as displacement vectors. 
The sign of the estimated \emph{informational} heat \(\Delta Q\) can be a tracked along a transition as an operational marker of reconfiguration (\(\Delta Q>0\)). The shaded “stabilized” and “high-repertoire” labels refer to relative differences in \(T\) and \(F_H\) within a dataset and are included only as descriptive guides; they are not universal thresholds nor claims about biological mechanism or equilibrium.

\subsection{Descriptions of experiments}

\subsubsection{Experimental paradigm: A-not-B EEG task}

The first dataset was obtained from a socially enriched version of the classic \textit{A-not-B} object-permanence task \cite{piaget2013construction,diamond1985development}. In this paradigm, an infant repeatedly retrieves a toy from location \textit{A}; when the toy is moved to location \textit{B}, younger infants often perseverate by re-reaching to \textit{A}, revealing limits in working memory and inhibitory control \cite{munakata1998infant}.  
EEG studies link successful performance to maturation of frontal networks and increased theta–alpha power \cite{bell1997individual, cuevas2011eeg}, while social cues from the caregiver modulate attention and dyadic neural synchrony \cite{feldman2007parent,reindl2018brain,santamaria2020emotional,leong2017speaker}.

Thirty-six mother–infant dyads (infants aged 12–24 months) completed up to eleven trials each, in which a toy was hidden under one of two cups. On half of the trials, the toy remained in the same location (\textit{non-switch}), and on the other half, it was moved to the opposite location (\textit{switch}). Each trial was coded as correct or incorrect based on the infant’s first reach, yielding four outcome types: (i) switch + correct, (ii) switch + error, (iii) non-switch + correct, and (iv) non-switch + error.

Simultaneous 32-channel dual-EEG was recorded from both mother and infant, allowing joint characterization of intra- and inter-brain dynamics. EEG windows (5 s) were converted into thermoinformational variables, entropy, internal energy, temperature, heat capacity, Helmholtz free energy, and heat, to quantify neural stability and information exchange during task performance.  

A detailed description of the experimental protocol, EEG setup, and preprocessing pipeline is provided in the Supplementary Section S2.1.

\subsubsection{Thermoinformational Characterization of EEG Dynamics in A-not-B Experiment}

Applying the thermoinformational framework to the same 5s EEG windows in order to derive physically interpretable state variables, entropy \(S\), internal energy \(U\), temperature \(T\), heat capacity \(C_{V}\), Helmholtz free energy \(F\), and  heat \(Q\).

\paragraph{Microstate ensemble.}  
Each window \(t\) of multi‐channel EEG was reduced via principal‐component analysis to an \(N\)-dimensional microstate vector  
$\mathbf{x}_{t}\in\mathbb{R}^{N}$. Collecting \(\{\mathbf{x}_{t}\}_{t=1}^{M}\) across all trials and participants defines an empirical ensemble from which to estimate a continuous density \(p(\mathbf{x})\).

\paragraph{Entropy options}
Alternative entropy formulations were evaluated to identify the most suitable description for each dataset.  
The Shannon entropy, quantifying the uncertainty of the empirical microstate distribution, was computed as  
\[
S \;=\; -\int p(\mathbf{x})\,\ln p(\mathbf{x})\,\mathrm{d}\mathbf{x},
\]
with \(p(\mathbf{x})\) estimated using Gaussian‐kernel density.  
To account for non‐Gaussian or heavy‐tailed structures, generalized entropies such as the Tsallis form \(S_q\) were also tested under the same numerical framework (see EEG application).  
The internal energy was defined analogously to the expectation of the microstate norm,
\[
U \;=\; \int \|\mathbf{x}\|\,p(\mathbf{x})\,\mathrm{d}\mathbf{x},
\]
ensuring thermodynamic consistency across entropy definitions.

\paragraph{Helmholtz free energy and Heat.}  
The Helmholtz free energy, as defined in \ref{eq:Helmholtz},
quantifies the stable, information bearing component of the system’s internal energy analogous to the usable work capacity in thermodynamics.  
To capture dynamic information exchange across time, the discrete heat increment between successive windows was defined as
\[
Q_{t\to t+1}
=\frac{T_{t}+T_{t+1}}{2}\,\bigl(S_{t+1}-S_{t}\bigr),
\]
and these increments were accumulated within each trial.

\paragraph{Interpretation.}  
Windows exhibiting lower $F$ correspond to more efficient or stable neural states, whereas peaks in $Q$ mark moments of rapid informational reorganization, often coinciding with task switches or errors. Higher values of $T$ and $S$ indicate operation in robust, high-repertoire regimes that precede correct choices rather than instability. In subsequent classification analyses, these thermoinformational variables outperform standard spectral measures in predicting infants’ A-not-B performance, underscoring the value of a physics grounded description of neural dynamics. 

\subsubsection{State-Space Density Computation}

To quantify how tightly the PCA-reduced microstates occupy their \(d\)-dimensional state space within each analysis window, we defined a scalar density
\[
\rho = \frac{n}{V},
\]
where \(n\) is the number of time samples (e.g., \(n=2500\)) and \(V\) is the effective volume of the minimum enclosing ellipsoid of the \(n\) PCA scores.  
The standardized \(n\times p\) data matrix \(X_{\mathrm{raw}}\) (zero mean and unit variance per channel) was projected onto its top \(d\) principal components:
\[
X = X_{\mathrm{raw,std}}\,P_{p\times d},
\]
yielding an \(n\times d\) representation. The empirical covariance matrix
\[
\Sigma = \frac{1}{n-1}\,X^\mathsf{T} X
\]
was then used to approximate the ellipsoid volume as
\[
V = (2\pi)^{d/2}\sqrt{\det(\Sigma) + \varepsilon},
\]
where a small \(\varepsilon > 0\) ensured numerical stability when \(\Sigma\) was nearly singular.  
Because \(n\) was constant across windows, fluctuations in \(\rho\) directly reflected geometric changes in the dispersion of the PCA-projected microstates: high \(\rho\) indicates tightly clustered (synchronous) states, whereas low \(\rho\) corresponds to more dispersed (desynchronized) dynamics.

\subsection{Optogenetic Stimulation Experiment}

The second dataset applies the thermoinformational framework to an optogenetically controlled social learning paradigm in mice, adapted from the classic Social Transmission of Food Preference (STFP) task \cite{galef1983transfer}.  
Mother–pup (\textit{dam–pup}) dyads underwent targeted optogenetic modulation of the medial prefrontal cortex (mPFC) to test whether neural synchrony influences social learning.  

Three conditions were compared: \textbf{Sync} (simultaneous 40 Hz stimulation of dam and pup), \textbf{Desync} (asynchronous stimulation with randomized phase lags), and \textbf{No-Opto} (unstimulated baseline).  In this work we are applying the methodology in order to identify the differences between the stimulated (Opto: Sync or Desync) and non-stimulated (No-Opto) mice.
Behavioral sessions comprised three phases (acquisition, interaction, and testing) recorded via multi-camera tracking. The durations of discrete social behaviors (approach, proximity, huddling) were treated as the system’s microstates, from which thermoinformational variables (entropy, internal energy, temperature, heat capacity, and Helmholtz free energy) were computed.

This experimental design enables direct comparison between the stability and information flux of social interaction under different synchronization regimes.  
A full description of animal preparation, viral strategy, stimulation parameters, and behavioral procedures is provided in the Supplementary Section S2.2.

\section{Results} 

\subsection{A-not-B Experiment Results}

In this experiment the EEG signal is used as a basis for the microstates. The formed microstate ought to have as many dimensions as the EEG channels to capture all the information. This leads to a very complex system. Due to the number of measurement per second in the EEG scarcity is not an issue, but in order to reduce complexity the dimensions need to be reduced.

The EEG was segmented into 5-second windows, and within each window Principal Component Analysis (PCA) was applied to reduce the 32-channel signal to its dominant components. For the main analysis, the top two principal components were retained, yielding a 2-dimensional state representation per window used to estimate the distribution functions for entropy and internal energy (Eq.~\ref{eq:Internal_Energy}). Entropy was computed using the Tsallis formulation with \(q=3\), as this non-extensive regime provided a better fit to the empirical EEG distributions. The probability densities of the principal components exhibited heavy tails consistent with a Student-\textit{t} rather than Gaussian profile (see Supplementary Section S5, Fig.~S9), supporting the use of a Tsallis-type entropy over the Shannon limit (\(q=1\)). The choice of \(q=3\) corresponds to the range of maximal sensitivity of the entropy measure to distributional changes in this heavy-tailed regime, while results remained stable for neighboring values of \(q\) within the non-extensive interval (\(2.75 \lesssim q \lesssim 3.25\)). Tests with 1--4 PCA components confirmed that 2--4 components produced comparable statistical separability across experimental conditions, whereas 1 component captured only approximately 40\% of the total variance and was insufficient to represent the system dynamics.

\begin{equation}
    \label{eq:Tsallis}
    S_q(p(E))=\frac{1}{q-1}\left(1-\int_{E_{min}}^{E_{max}}p(E)^qdE\right)
\end{equation}

Then based on the entropy and the internal energy the rest of the thermodynamic variables are calculated from Eqs. \ref{eq:1st_law},\ref{eq:Helmholtz}, and \ref{eq:Cv}.

\subsubsection{Modeling of outcomes: Toy switch and infant response}

\begin{figure}[htbp]
    \centering
    \includegraphics[width=1\linewidth]{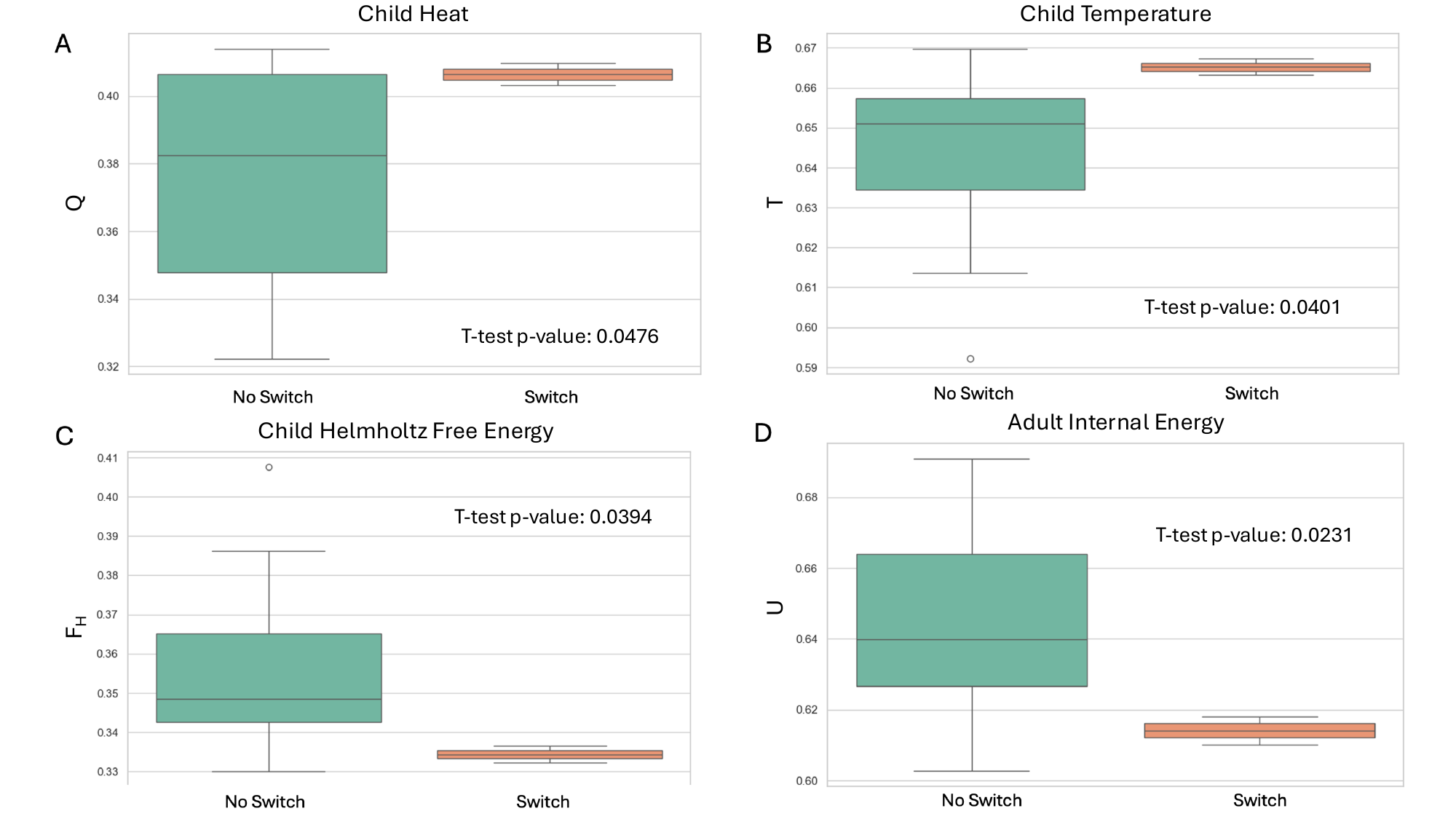}
    \caption{
    \textbf{Thermoinformational variables during toy \textit{switch} versus \textit{no-switch} trials.}
    (A) Child heat ($Q$) per pair, (B) child temperature ($T$), (C) child Helmholtz free energy ($F$), and (D)  adult internal energy ($U$)  all show significant differences between conditions ($p<0.05$). 
    Heat increases during \textit{switch} trials, marking stronger informational reconfiguration when the environment changes. 
    Temperature likewise increases, indicating a robust high-repertoire regime supporting adaptation. 
    Both free energy and adult internal energy decrease, consistent with more efficient and stable configurations after reorganization.
    Together, these results highlight a coherent thermodynamic pattern: \textit{high $Q$ and $T$ accompany reconfiguration}, whereas \textit{low $F$ and $U$ signal stabilization and efficiency.}
    }
    \label{fig:Switch_Predictions}
\end{figure}

\begin{figure}[htbp]
    \centering
    \includegraphics[width=1\linewidth]{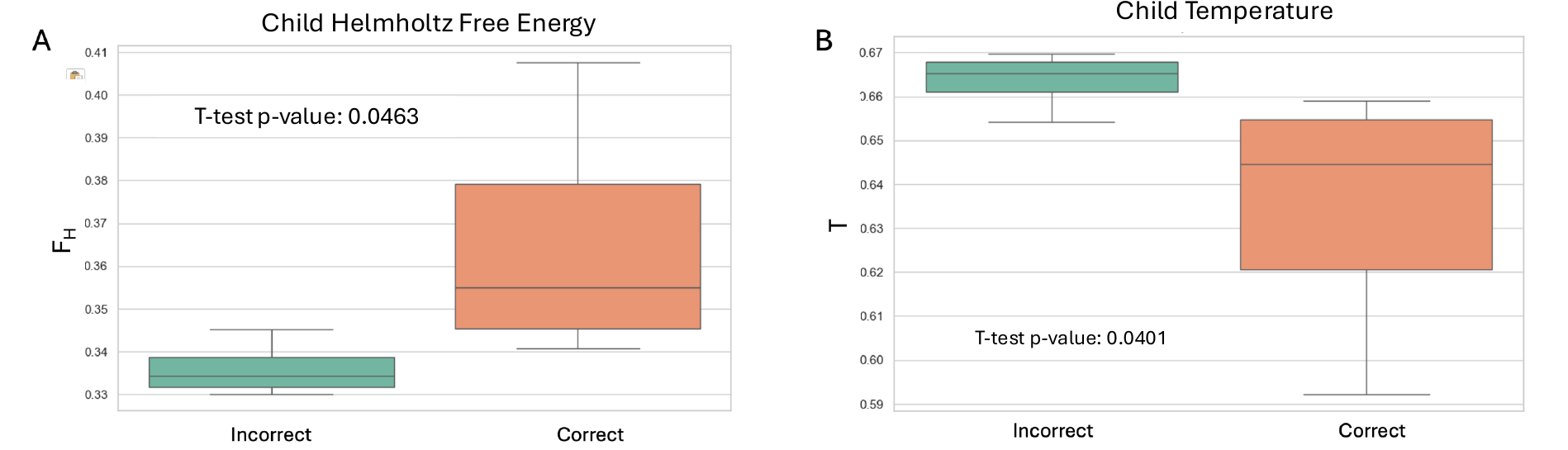}
    \caption{
    \textbf{Thermoinformational variables during correct versus incorrect responses.}
    (A) Child Helmholtz free energy ($F$) and (B) child temperature ($T$) show significant differences between correct and incorrect trials ($p<0.05$). 
    Correct choices are associated with slightly higher free energy and lower temperature, suggesting that successful retrieval involves transient increases in informational work capacity followed by efficient stabilization. 
    These complementary patterns reinforce the interpretation that thermoinformational variables differentiate \textit{phases of reconfiguration and stabilization} within adaptive neural behavior.
    }
    \label{fig:Answer_Predictions}
\end{figure}

Thermoinformational variables computed using the Tsallis entropy ($q=3$) with two principal components revealed consistent effects across both the \textit{switch} and \textit{answer} conditions (Figs.~\ref{fig:Switch_Predictions}–\ref{fig:Answer_Predictions}).

\paragraph{Switch condition.}
Four variables showed significant differences between \textit{switch} and \textit{no-switch} trials ($p<0.05$; Fig.~\ref{fig:Switch_Predictions}). 
Child heat ($Q$) increased during \textit{switch} trials, reflecting larger informational reconfiguration when the environment changed. 
Child temperature ($T$) also increased, indicating that during environmental updating the neural system occupies a robust high-repertoire regime, where additional activation produces minimal further entropy gain. 
In parallel, child Helmholtz free energy ($F$) decreased, consistent with a transition toward more efficient and stable configurations after adaptation. 
Adult internal energy ($U$) likewise decreased, suggesting reduced activation cost and greater energetic efficiency within the dyad during switching events. 
Together, these results indicate that thermoinformational variables capture both the energetic and entropic adjustments accompanying behavioral flexibility: high $Q$ and $T$ index reconfiguration, whereas low $F$ and $U$ mark stabilization and efficiency following adaptation. Specifically the statistical analysis revealed: for the Child Temperature Welch's t-test $p=0.0401$, $\Delta \mu = -0.0233$, $95\%$ CI $[-0.0453,-0.00137]$. For the Child Heat: Welch's t-test $p=0.0476$, $\Delta \mu = -0.0304$, $95\%$ CI $[-0.0605,-0.000409]$. For the Child Helmholtz Free energy: Welch's t-test $p=0.0394$, $\Delta \mu = 0.0235$, $95\%$ CI $[0.00147,0.0455]$. For the Adult Internal Energy: Welch's t-test $p=0.0231$, $\Delta \mu = 0.0316$, $95\%$ CI $[0.00562,0.0577]$. $\Delta \mu$ denotes the $NoSwitch - Switch$ difference in all the tests.

\paragraph{Answer condition.}
For the infant’s \textit{correct versus incorrect} responses (Fig.~\ref{fig:Answer_Predictions}), Helmholtz free energy ($F$) was modestly higher and temperature ($T$) slightly lower during correct choices ($p<0.05$). Specifically the statistical analysis revealed: for the Child Helmholtz Free energy Welch's t-test $p=0.0463$, $\Delta \mu = 0.0286$, $95\%$ CI $[0.000659,0.0566]$. For the Child Temperature: Welch's t-test $p=0.0401$, $\Delta \mu = -0.0284$, $95\%$ CI $[-0.0564,-0.000441]$. $\Delta \mu$ denoting the $Correct - Incorrect$ difference.
This suggests that successful retrieval engages transiently higher informational work capacity before settling into a stable state, consistent with the notion that correct performance requires controlled, energy-efficient reorganization rather than mere neural quiescence.

Overall, these findings demonstrate that the thermoinformational framework distinguishes between phases of neural reconfiguration (high $Q$, high $T$) and subsequent stabilization (low $F$, low $U$), providing a physically grounded description of how brain dynamics support adaptive behavior.

In addition to the individual interpretations of the thermoinformational variables, the same values can be used as inputs in a classifier to predict the different possible outcomes (correct or incorrect answer) and the different conditions (switch or not switch). The Child or Adult annotations describing if we are referring to the thermodynamic variable gotten from the adult's or the infant's data respectively.

\subsubsection{Classification of outcomes and conditions}

To evaluate the discriminative power of the thermoinformational variables, we trained simple classifiers to predict (i) the infant’s behavioral outcome (correct vs.\ incorrect) and (ii) the experimental condition (switch vs.\ no-switch).
Each model used \textit{two} thermoinformational features per task, selected for their physical interpretability (e.g., adult entropy paired with child Helmholtz free energy). 
Classification performance was evaluated using stratified five-fold cross-validation at the dyad level to prevent information leakage. 
Synthetic Minority Oversampling Technique (SMOTE) was applied within each training fold only, ensuring balanced classes without contaminating the test set. 
Reported metrics represent the mean (±SD) across folds, and the unit of analysis was the dyad-trial, with no dyad split between training and test folds.

\begin{table}[htbp]
\centering
\caption{Top-performing two-feature combinations for predicting behavioral outcomes and task conditions.}
\label{tab:Classification_TopPairs}
\small
\begin{tabular}{|l|l|c|c|c|c|}
\hline
\textbf{Task} & \textbf{Feature Pair} & \textbf{Accuracy} & \textbf{Precision} & \textbf{Recall} & \textbf{F1} \\ \hline
Answer (Correct vs.\ Incorrect) & Adult $S$, Child $F_H$ 
& 0.81 $\pm$ 0.14 & 0.92 $\pm$ 0.12 & 0.83 $\pm$ 0.24 & 0.84 $\pm$ 0.14 \\ \hline
Switch (Switch vs.\ No-switch) & Child $Q$, Diff $Q$  
& 0.97 $\pm$ 0.04 & 0.94 $\pm$ 0.08 & 1.00 $\pm$ 0.00 & 0.96 $\pm$ 0.05 \\ \hline
\end{tabular}
\end{table}

Across both behavioral outcome and condition classifications, compact two-dimensional representations of the thermoinformational state space yielded high and stable accuracies ($\approx 0.8$ for behavioral outcome, $\approx 0.97$ for condition). 
These results suggest that physically interpretable variables, entropy, temperature, internal energy, free energy, and heat, encode sufficient structure to distinguish behavioral and contextual states in EEG-derived data. 
Detailed results for all tested feature pairs and normalized confusion matrices are provided in Supplementary Section~S3 (Tables~S2–S5 and Fig.~S1).

\subsubsection{Comparison to the Free Energy Principle}

To benchmark the empirical Helmholtz free energy against Friston’s variational Free Energy Principle (FEP) \cite{friston2010free}, a minimal variational model was implemented on the same 5\,s EEG windows used in the thermoinformational analysis. The resulting variational free energy ($F_{\mathrm{FEP}}$) represents a bound on surprise within a specified generative model, whereas the thermoinformational free energy ($F_H = U - TS$) is derived directly from the empirical distribution of microstates, without requiring priors or latent variables. The full FEP model used and results are presented in the Supplementary Section S4.

Although both yield scalar “free energy” measures, they capture different but complementary aspects of brain dynamics. FEP formalizes the normative principle of Bayesian inference and evidence optimization, while $F_H$ provides a descriptive, data-driven quantity reflecting the physical stability and informational efficiency of neural activity. In the present data, 
$F_H$ was higher for correct than incorrect trials and lower for switch than no-switch trials, consistent with the respective stability interpretations, whereas $F_{\mathrm{FEP}}$ remained largely invariant, which may reflect the use of a basic FEP implementation in this application rather than a general property of FEP.

Taken together, the FEP and thermoinformational formulations can be viewed as complementary: the former offering a theoretical account of inference and adaptation, the latter providing a direct empirical characterization of the underlying thermodynamic organization of neural states.

\subsection{Optogenetic Synchronization Experiment Results}

In this experiment the behavioural microstates were defined as the durations of discrete social actions (approach, proximity, huddle). From the distribution of these durations we computed thermoinformational variables to summarise the stability and information flux of each dyad under Sync, Desync, or baseline conditions.

In this experiment the microstates used for the thermoinformational framework are the durations of the different actions. Based on these durations the thermoinformational variables of temperature, entropy, specific heat capacity, internal energy, and Helmholtz free energy. 

\subsubsection{Condition differentiation}

\textbf{Interaction}

\begin{figure}
    \centering
    \includegraphics[width=1\linewidth]{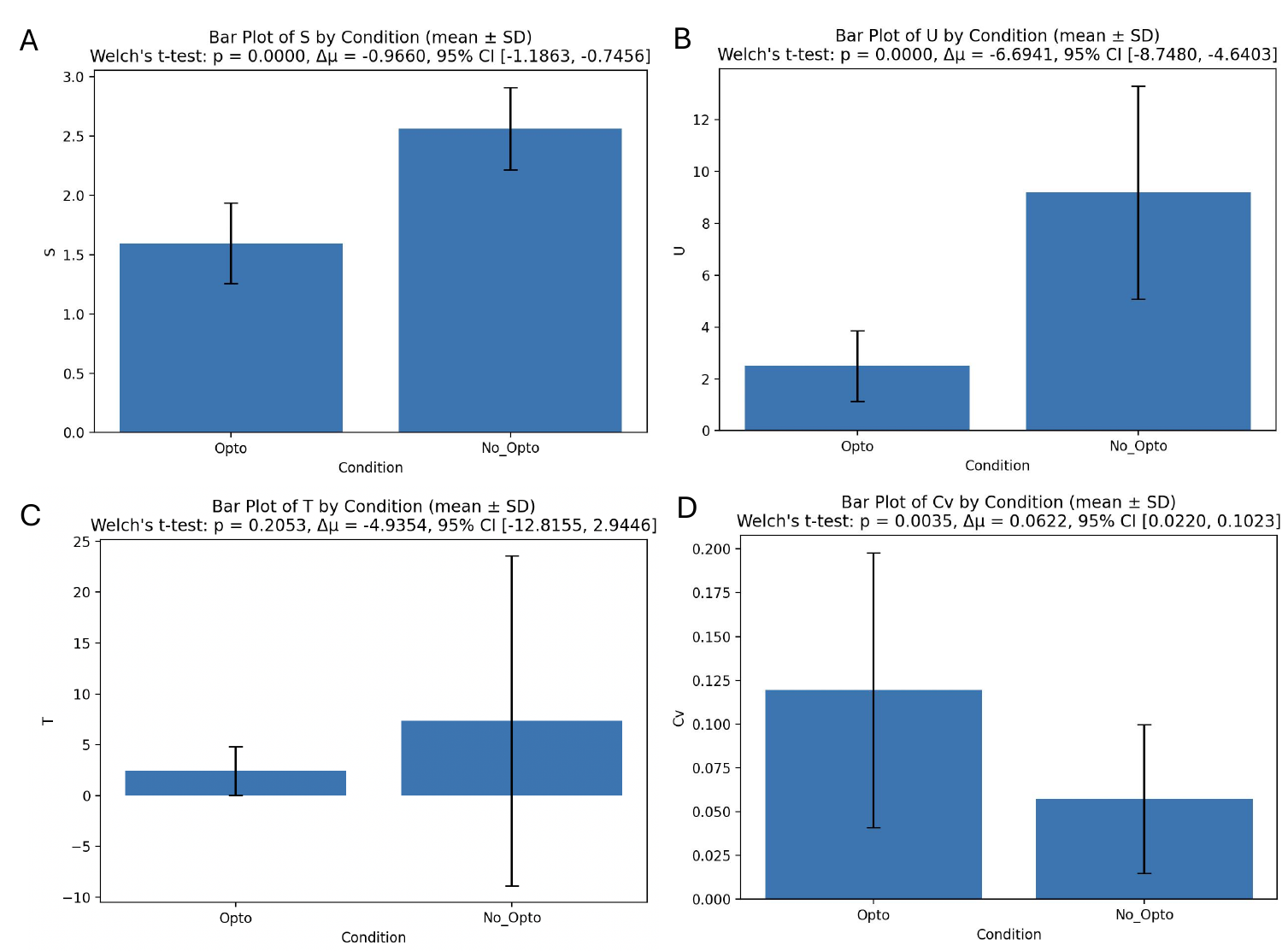}
     \caption{In this figure the mean values and standard deviations of the different thermoinformational variables are presented for the interaction phase. In subfigure A the entropy is displayed. The No Opto condition has higher entropy values with strong statistical significance. In subfigure B the internal energy is displayed, with the No Opto condition having higher values with a Welch's t-test  p-value $\le10^{-6}$. Subfigure C shows the temperature, with the No Opto condition having higher values but not statistically significant with a Welch's t-test  p-value of 0.2053. Lastly in subfigure D the Cv shows higher values for the Opto condition instead with a Welch's t-test p-value of 0.0035.}
    \label{fig:Interaction_Bar_Plot}
\end{figure}

In Fig. \ref{fig:Interaction_Bar_Plot}A the entropy of the Opto condition is seen to be smaller with a strong statistical difference. This means that the application of optogenetic synchrony on the mice leads to less varied behaviors, which lead to a smaller entropy. Similarly,  Figs. \ref{fig:Interaction_Bar_Plot}B and C depict that the Opto condition demonstrates smaller values. Given that the temperature is highly related to the variance of the microstate, this behavior also verifies that the Opto condition in the interaction phase indeed presents a more stable behavior. 

Lastly, in Fig. \ref{fig:Interaction_Bar_Plot}D it can be seen that the specific heat is higher in the Opto condition. This is also in accordance to the given interpretation. The specific heat capacity can be thought of as the "informational inertia" under these definitions. As such, it is expected that the more stable Opto condition also corresponds to a greater "informational inertia" of the system.

In the Supplementary Section S6, it is attempted to differentiate the Opto and No Opto conditions through their durations, and also through the more standard method of cRQA (cross-recurrence quantification analysis) commonly used in systems like this in experimental neuroscience. Both the cRQA metrics and the duration metrics, fail with almost all of their variables to pick up the different conditions with statistical difference. On the other hand, for the same phase all the thermoinformational variables are able to capture the condition difference with high statistical significance.

\textbf{Foraging}

\begin{figure}
    \centering
    \includegraphics[width=1\linewidth]{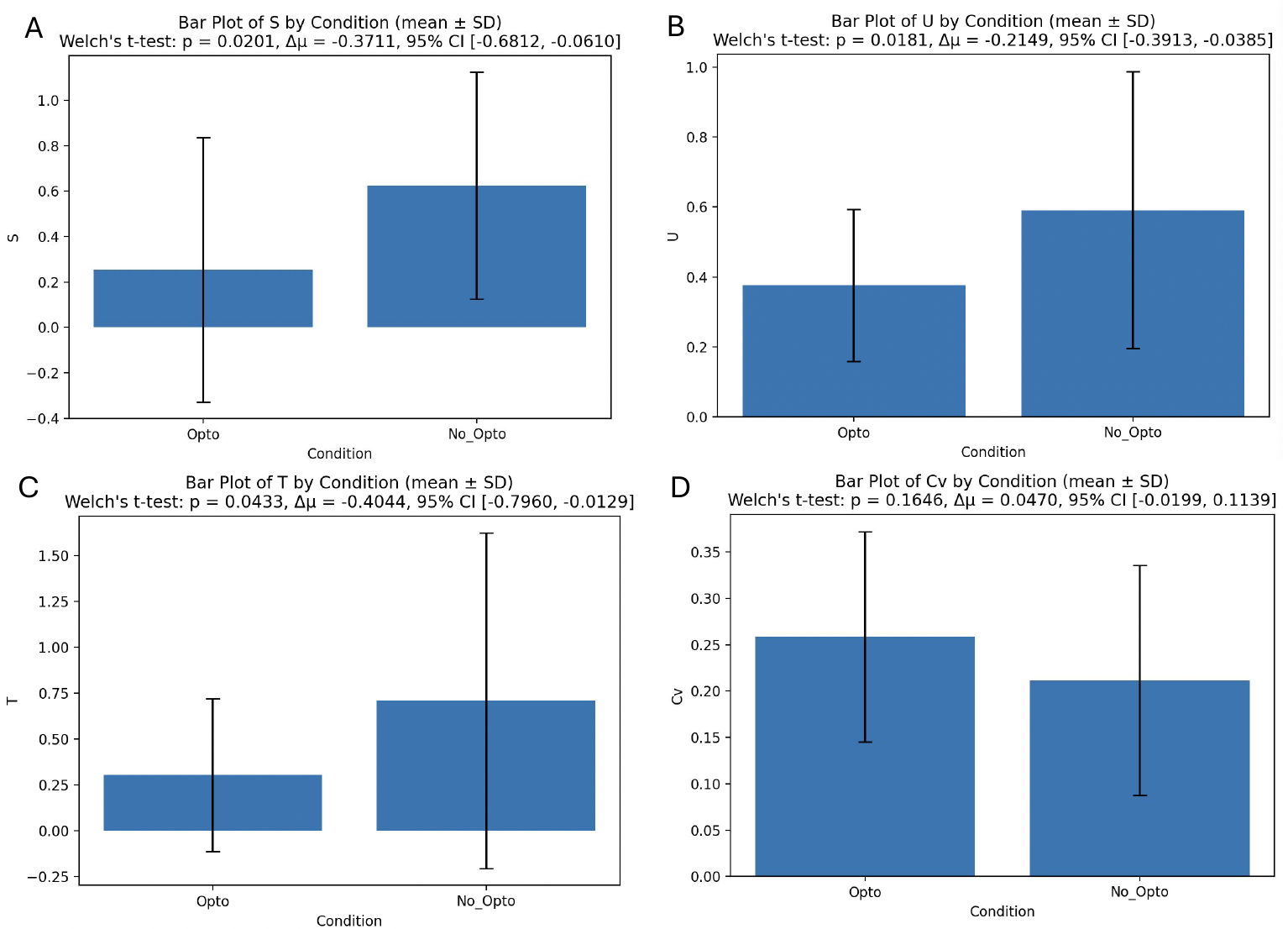}
    \caption{In this figure the mean values and standard deviations of the different thermoinformational variables are presented for the foraging phase. In subfigure A the entropy is displayed. The No Opto condition has higher entropy values with a Welch's t-test p-value 0.0201. In subfigure B the internal energy is displayed, with the No Opto condition having higher values with a Welch's t-test  p-value 0.0181. Subfigure C shows the temperature, with the No Opto condition having higher values with a Welch's t-test  p-value of 0.0433. Lastly in subfigure D while the Cv shows higher values for the Opto condition the difference is not statistically significant with a Welch's t-test p-value of 0.164}
    \label{fig:Foraging_Bar_Plots}
\end{figure}

Through subfigures A, B and C of Fig. \ref{fig:Foraging_Bar_Plots} it is observed that in the foraging stage as well the Opto condition offered greater stability than the no Opto condition. Given the fact that the optogenetic stimulation is only applied in the interaction phase, it is important that the difference in the thermoinformational variables is also retained across the phases of the experiment. This provides an important insight about the capability of these metric to capture the stability over time of a system. In  Supplementary Section S6, it is showcased how by studying just the behaviors of the durations themselves with a purely statistical approach it is not possible to capture this retained stability across the phases. 

As mentioned in the method section, in this type of thermoinformational framework it is possible to construct two types of variables, the energy based ones and the volumetric variables. These two types of variables in physics are derived from different properties of the same particles (energy and mass). In the presented case the duration of actions is the key variable used as an energy equivalent. In the foraging stage we can compute the volumetric variables as the ratio of food that the pup has eaten from each bowl. This is in line with the definition of the energy microstates as for each mouse now we would be using the time spent in a bowl as energy and the food eaten as mass. Specifically the density, which is to be used as the key volumetric variable,  will be defined as the ratio of food of the demo flavor (the flavor given to the mother mouse before the interaction stage) over the total food eaten by the pup in the foraging stage. Having defined the Helmholtz free energy Eq. \ref{eq:Helmholtz}, the temperature and the density, a fundamental equation of state of the form $F_H(\rho,T)$ can be constructed.

Using the values of these variable for each condition, a 3D state space representation is constructed. 3D splines are fitted to the data individually to the Opto ($R^2=0.8986$) and No Opto ($R^2=0.8758$) conditions.
\begin{figure}
    \centering
    \includegraphics[width=1\linewidth]{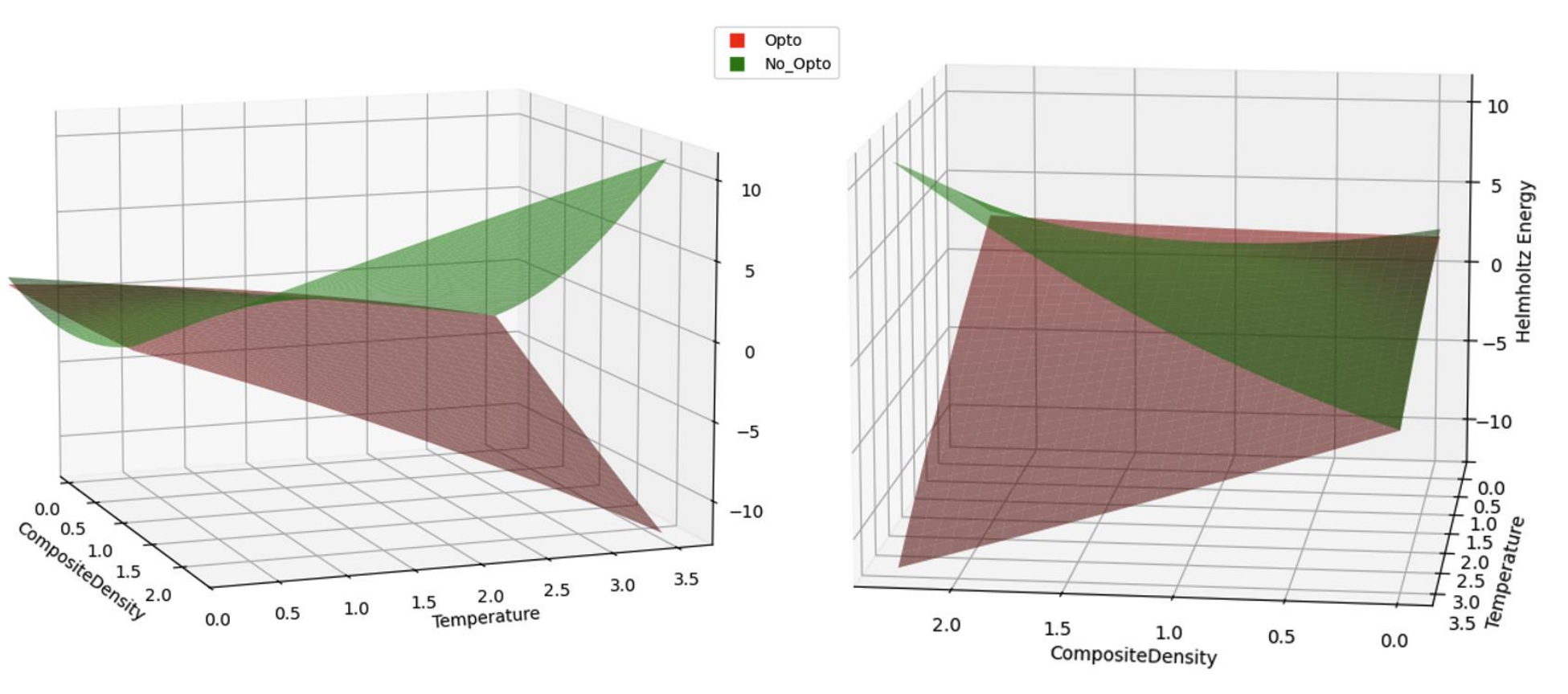}
    \caption{3D State space representation of the Foraging phase. The state space of the No Opto condition has a higher Helmholtz free energy than the Opto condition, signifying greater stability for the Opto condition}
    \label{fig:mice_3D_EoS}
\end{figure}

Figure \ref{fig:mice_3D_EoS} showcases that the Opto condition has a lower Helmholtz free energy compared to the No Opto in most of the 3D space. This indicates a more stable state with reduced drive to change; accordingly, the Opto condition occupies lower free energy regions. Thus, through this 3D representation it is further verified that the greater stability of the Opto condition is captured through the thermoinformational variables.

\paragraph{Comparison with conventional nonlinear metrics.}
To evaluate whether established time-series descriptors could capture the same condition differences, we compared the thermoinformational variables to a standard set of cross-recurrence quantification analysis (cRQA) measures (RR, DET, ENTR, LAM, etc.). 
Whereas cRQA metrics showed only weak or non-significant condition effects, all thermoinformational variables differentiated the optogenetic and baseline conditions with high statistical significance (see Supplementary Tables S1 and S2 in Supplementary Section S3). 
This suggests that the proposed framework captures information inaccessible to conventional dynamical metrics in this specific paradigm, offering a more physically grounded characterization of behavioral coordination.

\section{Conclusions}

The thermoinformational framework developed in this work provides quantitative interpretable summaries of the dynamics of two distinct neuroscientific systems. 
In the neuropsychological A-not-B EEG paradigm, thermoinformational variables derived directly from the signal reveal clear mechanistic signatures of cognitive adaptation. 
Trials in which the infant made a correct choice were characterized by higher Helmholtz free energy and lower temperature, indicating operation within a robust, high-repertoire regime that supports flexible yet stable decision-making.
Conversely, during environmental switches, the heat variable increased, reflecting stronger informational reconfiguration as the neural system adjusted to change. 
Together, these effects show that the thermoinformational state variables, entropy, temperature, free energy, and heat, capture both phases of neural behavior: reorganization under change and stabilization after adaptation. 

Moreover, these variables efficiently condense the information content of the system. 
When used as inputs to a simple classifier, they predict experimental outcomes and conditions with mean accuracies $\approx 80\%$ for behavioral outcome and higher for condition under dyad-level cross-validation with training-fold SMOTE, while retaining interpretability.
Previous approaches required far more complex machine-learning models to achieve lower performance and offered little mechanistic insight. 
In contrast, the thermoinformational framework achieves both precision and transparency, providing a unified physics-based description of brain and behavior that generalizes across the two different tested domains, indicating broader expandability in neuroscientific applications. 

\section*{Data and Software Availability}
All analysis code, figure-generation scripts, and environment files are publicly available at \newline 
\url{https://github.com/RDomenikos/Thermoinformatics_Neuroscience/tree/main}. To enable full computational reproducibility without exposing potentially re-identifiable signals, we publicly release de-identified, analysis-ready derivatives sufficient to reproduce all figures and statistics (feature matrices, PCA scores, and the computed S, U, T, F, and Q variables), together with detailed preprocessing documentation.
Because raw EEG may be re-identifiable, access to the raw recordings and trial-level metadata is provided under a controlled-access model governed by the NTU EMPOWER Board. Qualified researchers may request access via the repository landing page (see “Request Access”), agree to a data-use agreement prohibiting re-identification and onward sharing, and obtain approval from the EMPOWER Board. Upon approval, the data are delivered through NTU’s secure storage.

\section*{Funding}
This research is supported by RIE2025 Human Potential Programme Prenatal/Early Childhood Grants (H22P0M0002\&H24P2M0008), administered by A*STAR.

 \section*{Competing Interests}
The authors declare that they have no competing interests.

 \section*{Author Contributions}
G.R.D.: Conceptualization, Mathematical Modeling, Code, Statistical Analysis, Figure creation, Text Writing and Editing. V.L.: Supervision, Funding Acquisition. 

 \section*{Ethics}
All procedures involving human participants were approved by the Nanyang Technological University IRB. Written informed consent was obtained from all adult participants and from parents/guardians of infant participants prior to inclusion. All procedures involving animals were approved by the NTU Institutional Animal Care and Use Committee (IACUC protocol) and conformed to institutional and national guidelines.


\clearpage 

\nocite{*}

\bibliographystyle{sciencemag}
\bibliography{bibliography} 

%
%
%
%
%
%

\end{document}


\begin{center}
{\Large \textbf{Supplementary Materials for}}\\[4pt]
{\large \textit{A thermoinformational formulation for the description of neuropsychological systems}}\\[8pt]
{\normalsize George-Rafael Domenikos*, Victoria Leong}\\[12pt]
{\small$^\ast$Corresponding author. Email: georgios.rd@ntu.edu.sg}
\end{center}

This PDF file includes: \\[0.5em]
Supplementary Text: Sections S1 to S7\\
Figures: Fig. S1 to Fig. S7 \\
Tables: S1 to S3 \\

\section*{Supplementary Section S1. Example matrix representation of microstates}

For clarity, Table~S1 provides an illustrative example of how microstates are organized within the generalized data matrix \(\vec{A}\) described in the main text (Eq. 2).  
Each row corresponds to a set of measurements acquired at a given time (or other index), and each column represents a distinct measurement type for a specific individual.  
The matrix therefore defines the ensemble of possible microstates \(\vec{X}^t = (X_1^t, X_2^t, \dots, X_N^t)\), from which the probability distribution \(p(E)\) is constructed.

\begin{table}[htbp]
\centering
\caption*{\textbf{Table S1.} Example structure of the measurement matrix for a single individual.  
Each entry \(X_n^t|_i\) corresponds to the measurement of variable \(n\) at index \(t\) for individual \(i\).  
Rows correspond to different time points (or locations), and columns correspond to different measurement types.  
This structure generalizes naturally to higher-dimensional systems with multiple individuals or modalities.}
\label{tab:X_i}
\begin{tabular}{|l|l|l|l|l|l|l|}
\hline
$X|_i$ & $X_1$ & $X_2$ & . & . & . & $X_N$ \\ \hline
$t=0$  & $X^0_1|_i$ & $X^0_2|_i$ &   &   &   &   \\ \hline
$t=1$  & $X^1_1|_i$ &             &   &   &   &   \\ \hline
.      &             &             & . &   &   &   \\ \hline
.      &             &             &   & . &   &   \\ \hline
.      &             &             &   &   & . &   \\ \hline
$t=\tau$ &           &             &   &   &   & $X^\tau_N|_i$ \\ \hline
\end{tabular}
\end{table}

\noindent
In generalized form, the system matrix can be represented as
\[
\vec{A} = \vec{X}\{t, n, i\},
\]
where \(t = 0 \dots \tau\) indexes time (or location), \(n = 1 \dots N\) indexes the measurement types, and \(i = 1 \dots I\) indexes individuals.  
For systems of higher dimensionality, \(\vec{A}\) extends naturally to
\[
\vec{A} = \vec{X}\{a, b, \dots, \omega\},
\quad
\vec{A} \in \mathbb{R}^{a \times b \times \dots \omega},
\]
with the microstate defined as any \(\vec{x} \in \mathbb{R}^{(a \times b \times \dots \omega)-1}\).  
The dimensionality of the chosen microstate determines the type of information captured from the system.

\section*{Supplementary Section S2. Experiment Explanation}

\subsection*{S2.1 Experimental paradigm: A-not-B EEG task}
The first dataset derives from a socially enriched adaptation of the classic A‑not‑B object‑permanence task \cite{piaget2013construction, diamond1985development}.In its canonical form, an infant repeatedly retrieves a toy from location \textit{A}; when the toy is moved to location \textit{B}, many infants perseverate, re‑reaching to \textit{A}, revealing limits in working memory and inhibitory control \cite{munakata1998infant}. EEG work links successful performance to maturation of frontal networks and enhanced theta–alpha power \cite{bell1997individual, cuevas2011eeg}, and social cues from a caregiver modulate infants’ attention and dyadic neural synchrony \cite{feldman2007parent,reindl2018brain,santamaria2020emotional}.

Thirty‑six mother–infant dyads (infant age 12–24 months, $M=17.3±3.5months$) were recruited from the NTU Babylab volunteer panel.Infants were full‑term, typically developing, and free of neurological disorders; mothers reported no psychiatric diagnoses.All procedures were approved by the NTU Institutional Review Board, and written informed consent was obtained.

Both partners wore EasyCap 32‑channel caps connected to a dual BIOPAC Mobita wireless system arranged in the international 10‑10 layout \cite{santamaria2020emotional}. Signals were sampled at 500 Hz, referenced to Cz, and grounded at the nuchal midline to maximise infant comfort. Impedances were maintained $<10k\Omega$ for infants and $<20k\Omega$ for mothers. A hardware TTL trigger sent simultaneous pulses to both Mobita amplifiers and illuminated an LED captured on dual video streams, ensuring frame‑level alignment of EEG and behaviour. Data were band‑pass filtered (1–16 Hz), automatically rejected for $±80\mu V$ artefacts, and visually inspected (mean retention=83\% of epochs).

Dyads sat opposite each other at a rectangular table in a sound‑attenuated room. After a 60s joint baseline, each session comprised up to eleven trials (median = 10):

\begin{enumerate}
    \item \textbf{Attention cue} (2 s): experimenter rang a bell and called infant’s name.
    \item \textbf{Encoding}: mother lifted a soft toy, made eye contact for $\approx 1s$, then hid the toy in one of two opaque cups (locations\textit{A} or\textit{B}, 25cm apart).
    \item \textbf{Delay}: an occluder blocked view for 3s (standard working‑memory load\cite{diamond1985development, munakata1998infant}).
    \item \textbf{Search}: occluder removed; mother prompted, “Where’s teddy?”Infant’s first reach ($\le$10s) was coded online as correct/incorrect.
    \item \textbf{Outcome \& feedback}: toy revealed and brief neutral praise delivered to maintain engagement.
\end{enumerate}
Trials  1–2 always hid the toy at location\textit{A} to establish the prepotent response. Subsequent trials followed a pseudo‑random switch schedule: on half the trials the toy remained in the same cup (non‑switch), and on half it was transferred to the opposite cup while the infant watched (switch).This yields four outcome labels:

(i)switch + correct, (ii) switch + error (perseveration), (iii) non‑switch + correct, (iv) non‑switch + error (search failure). After artifact rejection and behavioral exclusions, 359 trials (mean 9.97 per infant) were retained.

This dual‑EEG implementation allows simultaneous analysis of intra‑individual brain dynamics and mother–infant neural coupling, variables implicated in cognitive flexibility and learning. The present dataset therefore provides a sound testbed for evaluating thermoinformational metrics of neural stability and information exchange.

\subsection*{S2.2 Experimental Paradigm: Optogenetic Synchronization Experiment}

The second experiment where the framework is applied to is the Optogenetic synchronization experiment on mice.

This experiment adapts the classic social transmission of food preference (STFP) task \cite{galef1983transfer} to mother–infant (dam–pup) dyads and combines it with bidirectional wireless optogenetic control of medial prefrontal cortex (mPFC) activity. The goal is to test whether intrabrain and interbrain neural synchrony modulate social learning.

\paragraph{Animals and viral strategy}

Wild‑type C57BL/6J dams and pups served as an unstimulated behavioural baseline, whereas the optogenetic groups used PV‑ChR2 mice generated by crossing PV‑Cre (B6; 129P2‑Pvalbtm1(cre)Arbr/J) with Ai32 (RCL‑ChR2(H134R) /EYFP) reporter mice for two generations. Final sample sizes were: Sync n=10 dyads, Desync n=13 dyads, after excluding two pups lacking opsin expression. Pups were tested at post‑natal day 23–29 (mean $25.6 ± 2.2$d); dams were 13–25 weeks old. Animals were maintained on a reverse 12h:12h light cycle and food‑restricted to 85–100\% baseline weight during testing. All procedures were approved by the NTU IACUC.

\paragraph{Surgery and wireless stimulation}

At post‑natal day 17–22 (pups) or 13–20 weeks (dams), mice were anaesthetised with isoflurane and implanted bilaterally with $250 \mu m$ optic fibres targeting mPFC (AP+1.9mm, ML±0.5mm, DV–1.5mm from pia).Fibres were affixed with dental cement, and animals recovered for $\ge7$d with analgesia. During behaviour, lightweight Teleopto receivers (2g dams; 1g pups) delivered 5ms blue pulses ($\approx 1–5mW$) in 2.5 min OFF–ON cycles under custom Bonsai/Arduino control \cite{lopes2015bonsai}. Two stimulation regimes were used:

\begin{itemize}
    \item \textbf{Sync} – identical 40 Hz trains to dam and pup (0 ms lag).
    \item \textbf{Desync} – pseudo‑random $40±1$Hz Gaussian trains independently phase‑shifted so that dam and pup pulses never overlapped within a 5ms window.
\end{itemize}
Wireless channels allowed independent control of each animal; histology confirmed fibre placement and ChR2 expression in PV interneurons, and slice recordings verified reliable $5–40$Hz light‑evoked firing.

\paragraph{Task structure}

The behavioural protocol comprised three phases executed in a sound‑ attenuating arena (30 × 30 × 35 cm) under red light:

\begin{enumerate}
    \item Acquisition (dam only, 60 min). Dams consumed 0.2g powdered chow flavored with either cinnamon (Cinn) or coriander (Cori).
    \item \textbf{Interaction (dam–pup, 30 min).} After a 5 min acclimation behind an opaque divider, dyads interacted through a perforated partition for 10 min (window sub‑phase) and then freely for 15 min. Optogenetic stimulation occurred only during the final 25 min. Social behaviors (approach, huddling, close‑proximity bouts) were recorded at 120 fps from four cameras and annotated frame‑by‑frame; these behavioral epochs constitute the microstates for downstream thermoinformational analysis.
    \item \textbf{Testing (pup only, 20 min).} Pups were placed in a two‑bowl arena containing Cinn and Cori chow. Video tracking quantified time in each bowl; bowl mass before/after yielded food consumed.
\end{enumerate}

The baseline condition is No-Opto where the mice are not subjects to any electric pulses. 

\section*{Supplementary Section S3. Classification of A-not-B Results}

\subsection*{S3.1 Classification Predictions}

Classification performance was evaluated using stratified five-fold cross-validation at the dyad level (32 dyads, up to 11 trials per dyad), ensuring that all trials from a given dyad were contained within a single fold to prevent information leakage. Each reported accuracy, precision, recall, and F1 score represents the mean across folds, with standard deviations reflecting variability across folds. All classifiers were implemented as Random Forests without hyperparameter tuning. Synthetic Minority Oversampling Technique (SMOTE) was applied *within each training fold only* to balance class representation without contaminating test data. The unit of analysis is the dyad-trial, not the individual time window.

For interpretability and physical grounding, we restricted inputs to combinations of **two thermoinformational variables**, each corresponding to meaningful physical quantities such as entropy, temperature, internal energy, Helmholtz free energy, and heat, or their adult–child differences. To illustrate classifier behavior, average normalized confusion matrices for both tasks are provided in Supplementary Fig.~S1.

\subsubsection*{S3.1.1 Answer Predictions (Tables~S2–S3)}

The classifier predicted the infant’s behavioral outcome (correct vs.\ incorrect) in the A-not-B experiment using two-variable combinations of thermoinformational descriptors. The best-performing pairs achieved accuracies around 0.80, demonstrating that compact thermodynamic representations carry sufficient information to predict behavioral success.

\begin{table}[h!]
\centering
\caption*{\textbf{Table S2.} Mean performance metrics (±SD) for answer prediction using the top feature pairs.}
\begin{tabular}{|l|c|c|c|c|}
\hline
\textbf{Feature Pair} & \textbf{Accuracy} & \textbf{Precision} & \textbf{Recall} & \textbf{F1} \\ \hline
Adult S, Child F & 0.81 $\pm$ 0.14 & 0.92 $\pm$ 0.12 & 0.83 $\pm$ 0.24 & 0.84 $\pm$ 0.14 \\ \hline
Adult S, Child T & 0.81 $\pm$ 0.14 & 0.92 $\pm$ 0.12 & 0.83 $\pm$ 0.24 & 0.84 $\pm$ 0.14 \\ \hline
Adult S, Adult Q & 0.72 $\pm$ 0.04 & 0.72 $\pm$ 0.04 & 1.00 $\pm$ 0.00 & 0.84 $\pm$ 0.03 \\ \hline
\end{tabular}
\end{table}

\begin{table}[h!]
\centering
\caption*{\textbf{Table S3.} Average normalized confusion matrix for the answer prediction task.}
\begin{tabular}{|c|c|c|}
\hline
 & \textbf{Predicted Correct} & \textbf{Predicted Incorrect} \\ \hline
\textbf{Actual Correct} & 0.79 & 0.21 \\ \hline
\textbf{Actual Incorrect} & 0.15 & 0.85 \\ \hline
\end{tabular}
\end{table}

\subsubsection*{S3.1.2 Switch Predictions (Tables~S4–S5)}

For the condition classification (switch vs.\ no-switch), results were obtained using two-variable feature pairs. The model achieved higher accuracies (up to 0.97), indicating that thermoinformational measures are particularly sensitive to condition-dependent changes in EEG-derived state dynamics. SMOTE was applied within training folds to balance the smaller number of switch trials.

\begin{table}[h!]
\centering
\caption*{\textbf{Table S4.} Mean performance metrics (±SD) for switch condition prediction using the top feature pairs.}
\begin{tabular}{|l|c|c|c|c|}
\hline
\textbf{Feature Pair} & \textbf{Accuracy} & \textbf{Precision} & \textbf{Recall} & \textbf{F1} \\ \hline
Child Q, Diff Q & 0.97 $\pm$ 0.04 & 0.94 $\pm$ 0.08 & 1.00 $\pm$ 0.00 & 0.96 $\pm$ 0.05 \\ \hline
Child Q, Diff F & 0.94 $\pm$ 0.05 & 0.90 $\pm$ 0.07 & 1.00 $\pm$ 0.00 & 0.93 $\pm$ 0.05 \\ \hline
Child F, Child Q & 0.94 $\pm$ 0.04 & 0.89 $\pm$ 0.08 & 1.00 $\pm$ 0.00 & 0.93 $\pm$ 0.05 \\ \hline
Child T, Child Q & 0.94 $\pm$ 0.04 & 0.89 $\pm$ 0.08 & 1.00 $\pm$ 0.00 & 0.93 $\pm$ 0.05 \\ \hline
Child Q, Diff S & 0.95 $\pm$ 0.04 & 0.88 $\pm$ 0.10 & 0.92 $\pm$ 0.12 & 0.89 $\pm$ 0.10 \\ \hline
\end{tabular}
\end{table}

\begin{table}[h!]
\centering
\caption*{\textbf{Table S5.} Average normalized confusion matrix for switch condition prediction.}
\begin{tabular}{|c|c|c|}
\hline
 & \textbf{Predicted No-Switch} & \textbf{Predicted Switch} \\ \hline
\textbf{Actual No-Switch} & 0.78 & 0.22 \\ \hline
\textbf{Actual Switch} & 0.00 & 1.00 \\ \hline
\end{tabular}
\end{table}

\vspace{6pt}
\noindent
Across both behavioral outcome and condition classifications, thermoinformational variables provided stable and interpretable predictive performance using only two features. The results demonstrate that the thermodynamic descriptors (entropy, temperature, free energy, internal energy, and heat) capture sufficient structure in the EEG-derived state space to distinguish both behavioral success and task condition with high accuracy under rigorous cross-validation and class-balancing controls.

---

\section*{Supplementary Section S4. Comparison to the Free Energy Principle}

\subsection*{S4.1 Application of the Variational Free-Energy Principle to A-not-B EEG}
The idea of utilising a free energy framework in order to model neuroscientific systems was first established by the ground breaking work of Karl Friston \cite{friston2010free}. 
To benchmark our data-driven Helmholtz free-energy against Friston’s Free-Energy Principle (FEP), we implemented a minimal variational model on each 5s EEG window (2500 samples) to obtain a trial-wise variational free-energy \(F_{\mathrm{FEP},t}\).  Our procedure comprised four stages: feature extraction, generative model specification, parameter estimation, and variational inference.

Each 5s segment of infant (and adult) EEG was first mean-imputed column-wise, then reduced via PCA.  Let
\[
y_{t}\in\mathbb{R}^{N}
\;=\;\frac{1}{\tau_{tot}}\sum_{\tau=1}^{\tau_{tot}}\mathrm{PCA}_{1:N}\bigl(\mathrm{EEG}(\tau)\bigr),
\quad
\tau_{tot}=2500,\;N=2
\]
denote the mean of the first \(N\) PC scores over the window.

We introduced a binary latent state \(s_{t}\in\{0,1\}\) per window, interpreted as the unobserved “correct-choice belief.”  The joint density was
\[
\begin{aligned}
p(s_{t}\mid \mathrm{Switch}_{t})
&=\sigma\bigl(\alpha\,\mathrm{Switch}_{t} + \beta\bigr),
\quad
\mathrm{Switch}_{t}\in\{0,1\},\\
p(y_{t}\mid s_{t})
&=\mathcal{N}\bigl(y_{t}\mid W\,s_{t},\,\Sigma\bigr),
\end{aligned}
\]
where \(\sigma(x)=(1+e^{-x})^{-1}\), \(W\in\mathbb{R}^{N}\), and \(\Sigma\in\mathbb{R}^{N\times N}\).

Pooling all windows across trials and participants, labeled by observed outcome \(y^{\mathrm{obs}}\in\{0,1\}\) (error vs.\ correct), we set
\[
W \;=\;\mathbb{E}[\,y_{t}\mid y^{\mathrm{obs}}=1\,]\;-\;\mathbb{E}[\,y_{t}\mid y^{\mathrm{obs}}=0\,],
\qquad
\Sigma \;=\;\mathrm{Cov}\bigl(y_{t} - W\,s^{\mathrm{obs}}_{t}\bigr) \;+\;\epsilon I,
\]
with \(\epsilon=10^{-6}\).  The prior parameters \((\alpha,\beta)\) were fit by logistic regression predicting \(y^{\mathrm{obs}}\) from \(\mathrm{Switch}_{t}\).

Assuming \(q(s_{t})=\mathrm{Bernoulli}(\mu_{t})\), the posterior mean update is
\[
\mu_{t}
=\sigma\!\Bigl(\alpha\,\mathrm{Switch}_{t} + \beta
\;+\;W^{\mathsf{T}}\Sigma^{-1}y_{t}
\;-\tfrac12\,W^{\mathsf{T}}\Sigma^{-1}W\Bigr).
\]
The trial-wise variational free energy is
\[
\begin{split}
F_{\mathrm{FEP},t}
&=\mathbb{E}_{q}\bigl[\ln p(y_{t},s_{t})\bigr]
\;-\;\mathbb{E}_{q}\bigl[\ln q(s_{t})\bigr]\\
&=\mu_{t}\ln\mathcal{N}(y_{t}\mid W,\Sigma)
\;+\;(1-\mu_{t})\ln\mathcal{N}(y_{t}\mid 0,\Sigma)
\;+\;\bigl[-\mu_{t}\ln\mu_{t}-(1-\mu_{t})\ln(1-\mu_{t})\bigr].
\end{split}
\]
Computing \(F_{\mathrm{FEP},t}\) for every 5s window in both infant and adult EEG provided scalar time series \(\{F_{\mathrm{FEP},t}^{\mathrm{inf}}\}\) and \(\{F_{\mathrm{FEP},t}^{\mathrm{adm}}\}\).

\subsection*{S4.2 Comparison of the Variational Free Energy to the data driven presented Framework, Figs. S1-S2}

\begin{figure}
    \centering
    \includegraphics[width=1\linewidth]{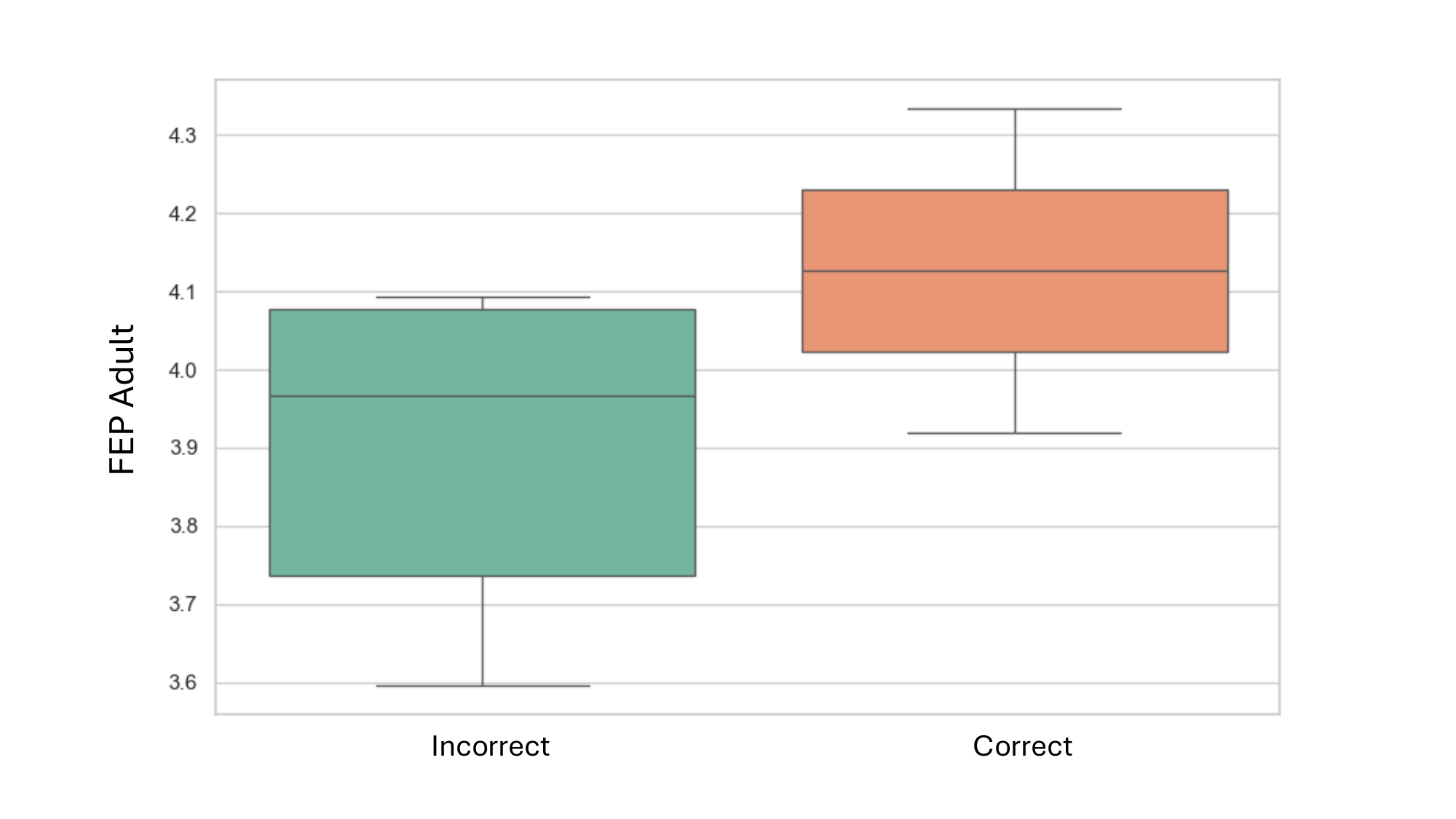}
    \caption*{\textbf{Fig. S1.} Mean Helmholtz Free Energy of the mother's EEG  per trial comparison for each condition for correct and incorrect answer of the child, with a t-test $p=0.467$. }
    \label{fig:FEP_adult}
\end{figure}

\begin{figure}
    \centering
    \includegraphics[width=1\linewidth]{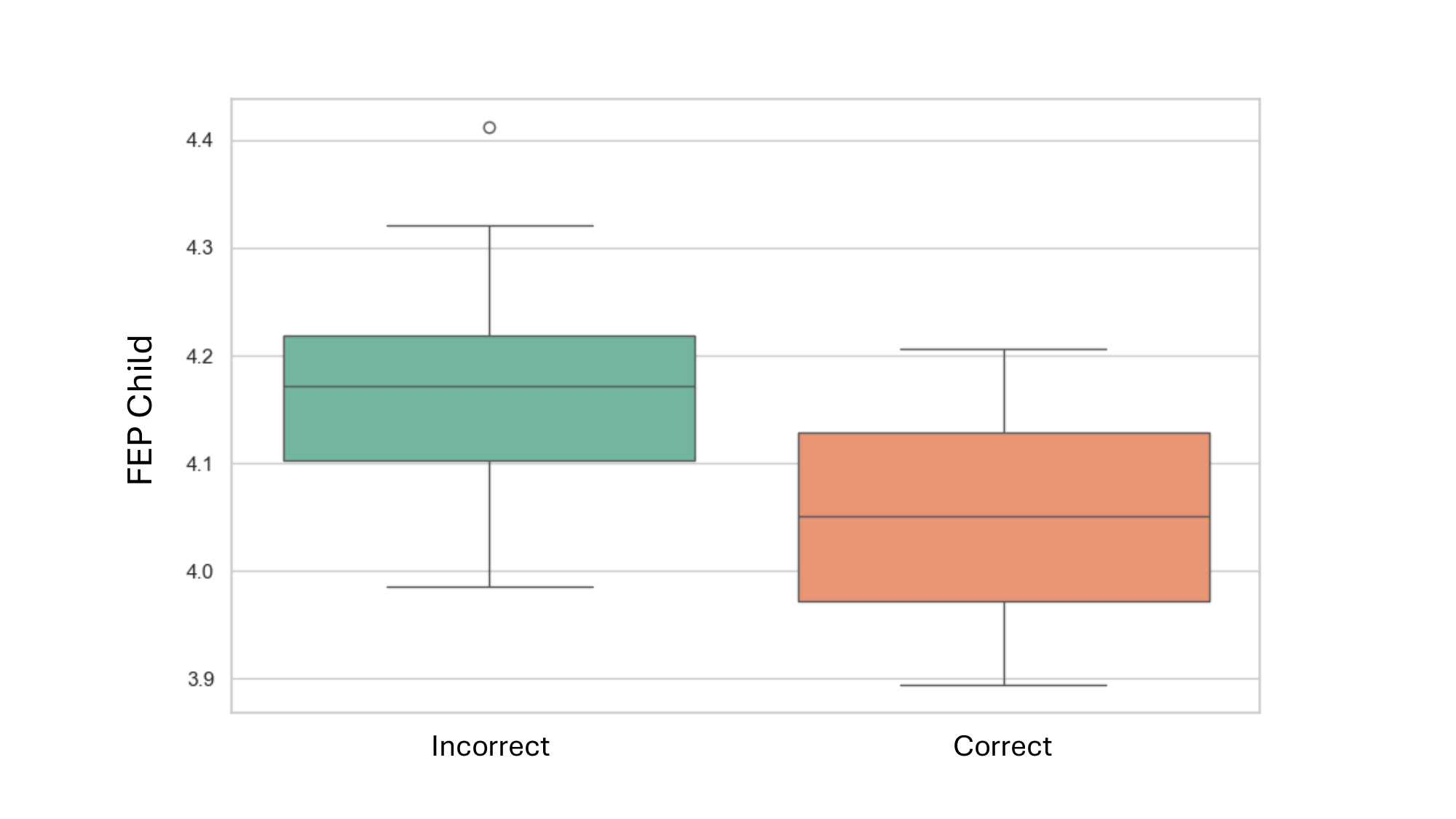}
    \caption*{\textbf{Fig. S2.} Mean Helmholtz Free Energy of the child's EEG  per trial comparison for each condition for correct and incorrect answer of the child, with a t-test $p=0.576$.}
    \label{fig:FEP_child}
\end{figure}

Although both our thermoinformational Helmholtz free energy and Friston’s variational free‐energy yield a scalar “free‐energy” measure for each EEG window, they diverge in fundamental ways.  The FEP is inherently model‐dependent, requiring an explicit generative model with latent states and priors, here a Bernoulli‐Gaussian formulation tuned to task structure, so that its free‐energy quantifies surprise under that specific inference scheme.  By contrast, the Helmholtz free energy of the thermoinformational model is computed directly from the empirical distribution of microstates (via Monte Carlo density estimation, PCA, and spline‐based derivatives) without latent variables or priors, making it fully data‐driven and portable across domains.  Moreover, FEP’s reliance on variational Bayesian updates entails iterative optimization and Gaussian‐integral computations, whereas the thermoinformational pipeline scales naturally through GPU‐accelerated kernel‐density and principal‐component routines without costly E–M loops.  Finally, the thermoinformational variables—entropy, temperature, heat capacity, and free energy—admit immediate interpretations as measures of neural disorder, stability, and usable information, whereas \(F_{\mathrm{FEP}}\) reflects an evidence bound tied to the chosen model architecture.  Together, these differences underscore that Friston’s FEP and the thermoinformational framework offer complementary perspectives: one grounded in Bayesian belief updating, the other in model‐agnostic physical state variables that capture the intrinsic thermodynamics of neural activity.

In some key aspects the thermoinformational description is able to provide certain advantages over the FEP due to inherent  advantages in its nature. First, by directly estimating the full microstate distribution via Monte Carlo KDE rather than relying on simple variational approximations (e.g.\ Gaussian recognition densities), it achieves tighter data-driven estimates that are sensitive to subtle, non-Gaussian features in the dynamics. Second, whereas the FEP merges entropy and energy into a single free-energy term, masking the individual contributions of model complexity and data fit, thermoinformatics explicitly computes Shannon (or Tsallis) entropy and internal energy as separate quantities and then uses the first law of thermodynamics to ensure proper additivity; this means that two conditions with identical variational bounds under FEP can nonetheless be distinguished by differences in temperature or internal energy in our framework. Third, the FEP typically treats each EEG channel or trial in isolation, whereas thermoinformatics captures the true “information flux” between states via heat calculations, revealing dyadic interactions that were not evident through a basic FEP implementation. Finally, because FEP generally assumes a slow relaxation toward a steady-state attractor, it cannot account for momentary, inversions of the second law, brief, non-equilibrium excursions that increase order, whereas our windowed, assumption-free pipeline can detect short, non-equilibrium excursions in the windowed analysis.

---

\section*{Supplementary Section S6. Benchmark Comparisons of Thermoinformational variables vs cRQA and Durations, Figs S3-S4}

\begin{figure}[h!]
\centering
\includegraphics[width=1\linewidth]{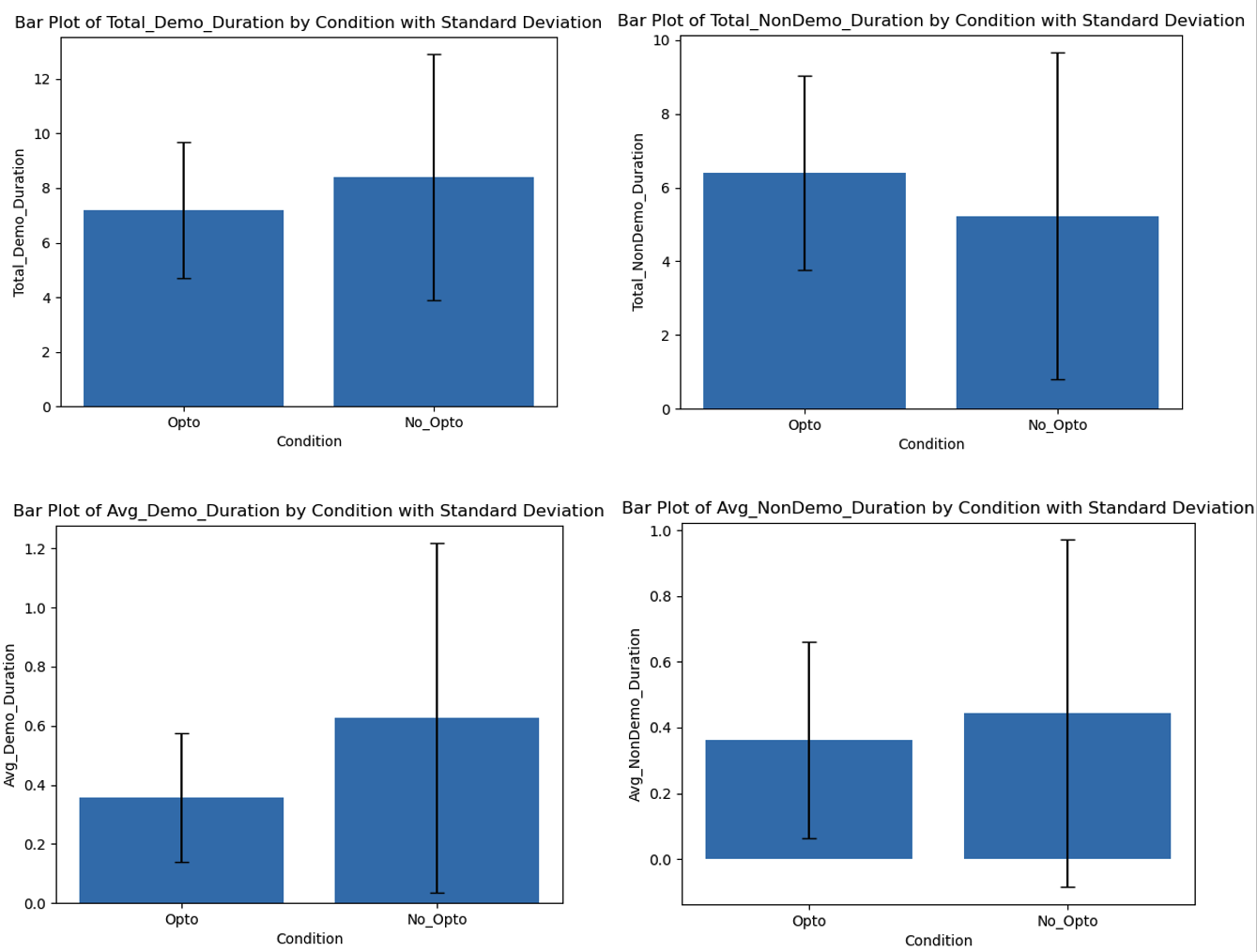}
\caption*{\textbf{Fig. S3.} Comparison using duration-based metrics. None of the duration variables reached statistical significance between Opto and No-Opto groups (Total Demo: $p=0.2517$; Total NonDemo: $p=0.2620$; Avg Demo: $p=0.043$; Avg NonDemo: $p=0.5063$). In contrast, thermoinformational variables in Fig.~4 of the main text showed highly significant effects.}
\end{figure}

\begin{figure}[h!]
\centering
\includegraphics[width=1\linewidth]{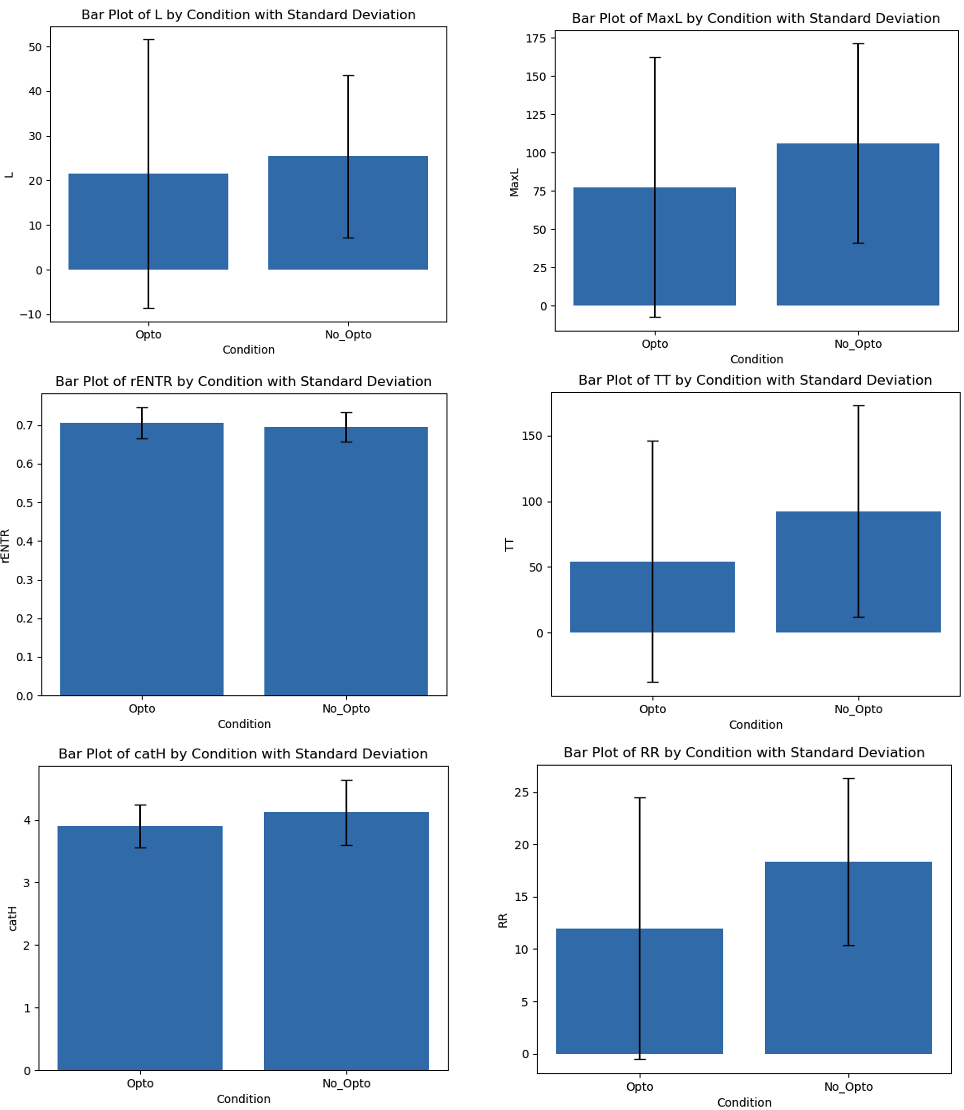}
\caption*{\textbf{Fig. S4.} Comparison using cRQA metrics. None of the standard nonlinear measures discriminated between Opto and No-Opto conditions (L: $p=0.547$; MaxL: $p=0.1585$; TT: $p=0.107$; rENTR: $p=0.3069$; catH: $p=0.086$; RR: $p=0.0255$).}
\end{figure}

\section*{Supplementary Section S5: Empirical justification for non-extensive entropy (Tsallis \(q = 3\)), Fig. S5}

The entropy used in the thermoinformational analysis was computed using the Tsallis non-extensive formulation with \(q = 3\), applied to the probability distribution of EEG microstates.  
To obtain the empirical microstate ensemble, each 5\,s EEG window was projected onto its first two principal components (\(N = 2\)), representing the dominant modes of neural activity while preserving the geometry of the state space.  
The joint distribution \(p(x_{\mathrm{PC1}}, x_{\mathrm{PC2}})\) of these PCA-reduced signals revealed pronounced leptokurtic structure with extended tails, deviating from a Gaussian profile.

\begin{figure}[htbp]
    \centering
    \label{fig:S_Gaussian}
    \includegraphics[width=0.85\linewidth]{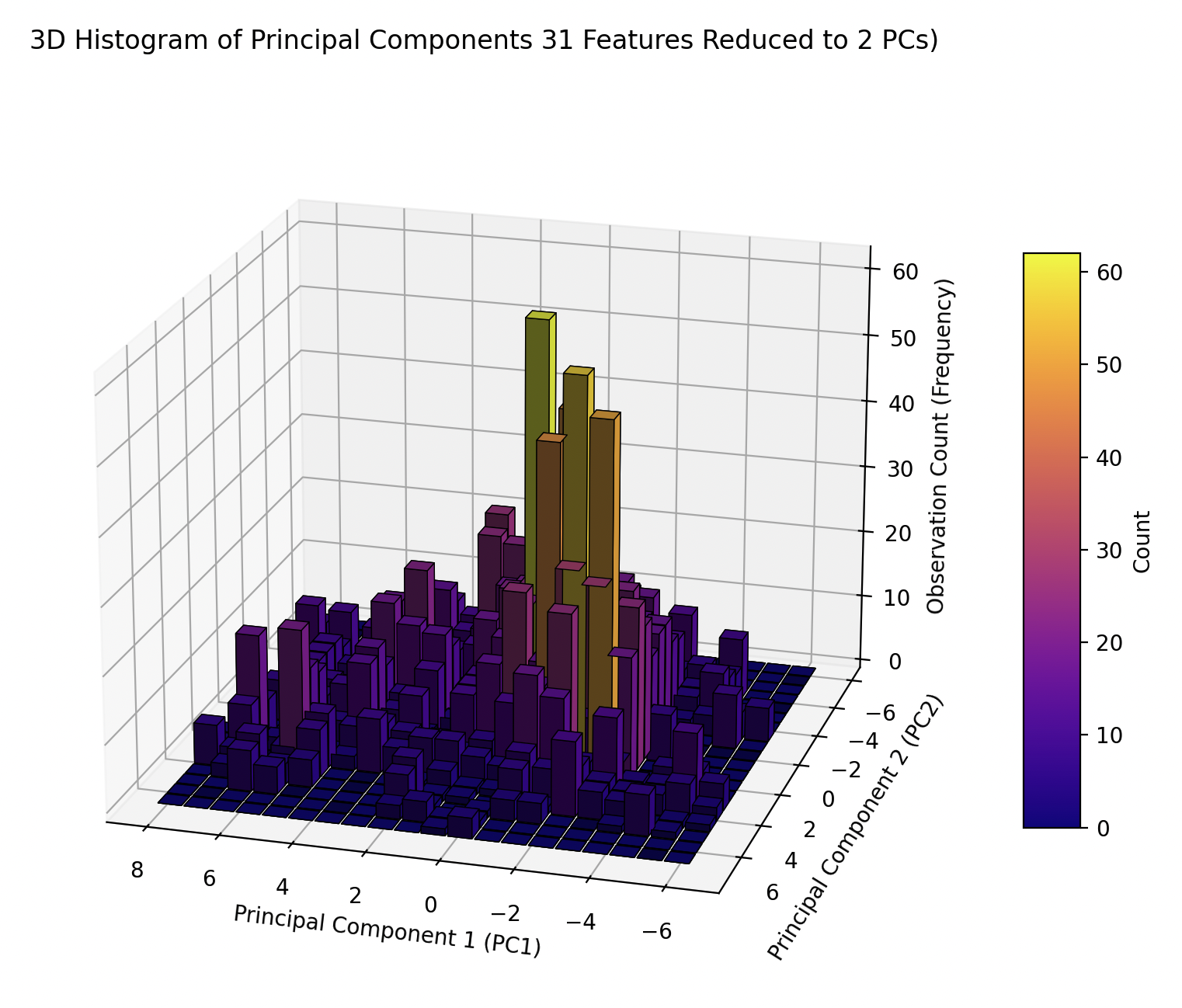}
    \caption*{\textbf{Fig.\,S5.} 
    3D histogram of PCA-reduced EEG microstates (31 channels reduced to two principal components).  
    The heavy-tailed, anisotropic distribution demonstrates clear departure from Gaussianity, with broad probability mass extending along both principal axes.  
    Such non-Gaussian statistics justify the use of the Tsallis entropy with \(q = 3\), which accurately captures non-extensive, heavy-tailed fluctuations beyond the Shannon limit (\(q \to 1\)).}
\end{figure}

These observations confirm that the EEG microstate ensemble occupies a non-extensive statistical regime, consistent with prior reports of power-law and Student-\(t\) (as seen in Fig. S5 distributed activity in cortical dynamics.  
Accordingly, the thermoinformational variables derived from this entropy, \(S_q\), \(T_q\), and \(F_q\), better describe the true variability, stability, and information flux of the neural system than additive entropies based on Gaussian assumptions.

\section*{Supplementary Section S7. Correlation of Interaction and Foraging Stages in Optogenetic Experiment, Fig. S6-S7}

\begin{figure}[h]
\centering
\includegraphics[width=1\linewidth]{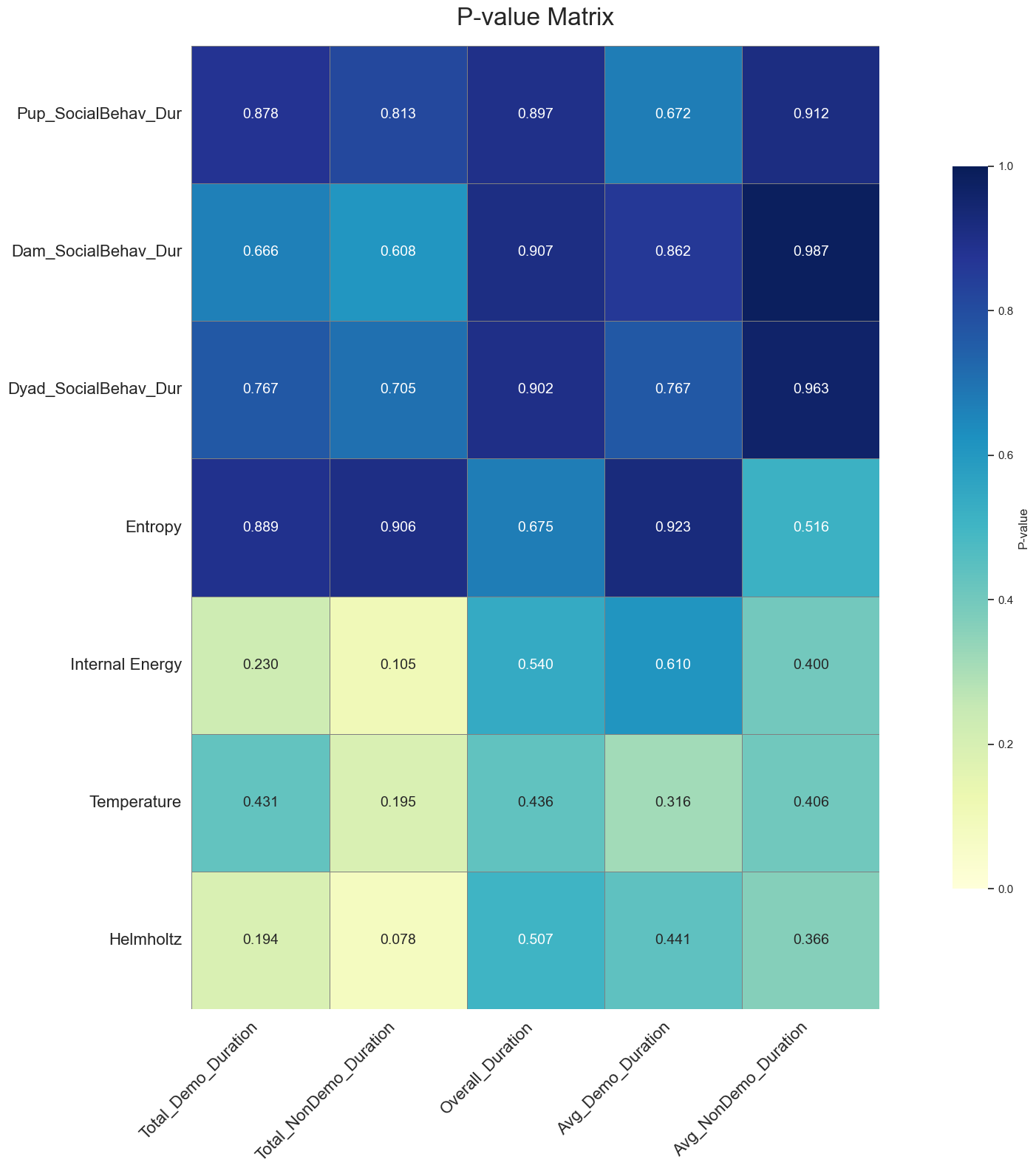}
\caption*{\textbf{Fig. S6.} Correlations between interaction-phase metrics (duration and thermoinformational) and foraging-phase durations, shown via corresponding $p$-values.}
\end{figure}

\begin{figure}[h]
\centering
\includegraphics[width=0.75\linewidth]{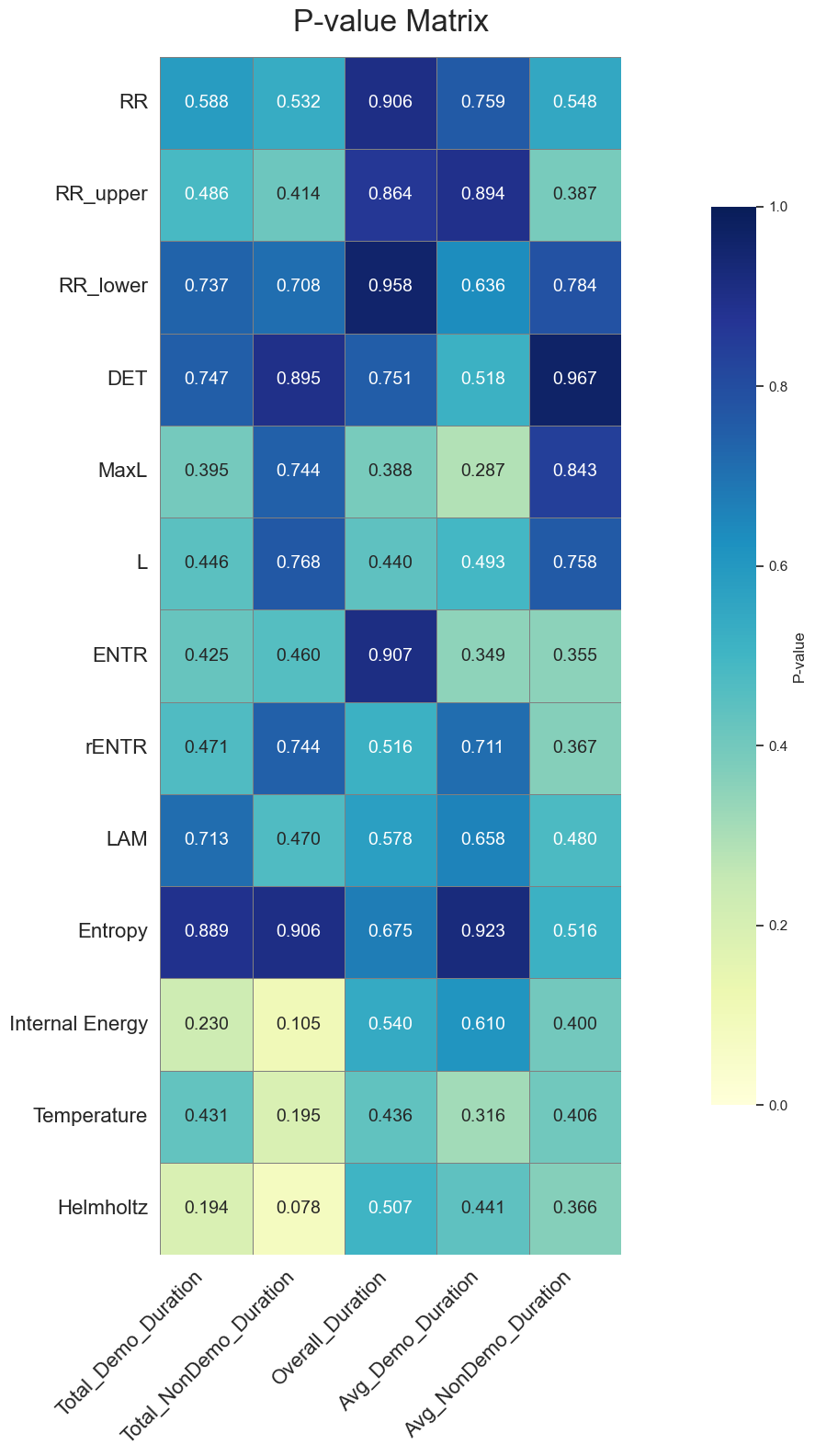}
\caption*{\textbf{Fig. S7.} Correlations between interaction-phase metrics (cRQA and thermoinformational) and foraging-phase durations, expressed as $p$-values.}
\end{figure}

\bibliographystyle{sciencemag}